\documentclass{aa}  
\usepackage[switch]{lineno}
\nolinenumbers
\usepackage{lastpage}
\usepackage{graphicx}
\usepackage{float}
\usepackage{txfonts}

\usepackage[]{hyperref}
\hypersetup{
    colorlinks=true,    
    linkcolor=blue,     
    citecolor=blue,     
    filecolor=magenta,  
    urlcolor=blue       
}
\makeatletter
\renewcommand*\aa@pageof{, page \thepage{} of \pageref*{LastPage}}
\makeatother
\usepackage{natbib}
\bibpunct{(}{)}{;}{a}{}{,} 
\usepackage{booktabs}
\usepackage{pdflscape}
\usepackage{multirow}
\usepackage{threeparttable}
\usepackage{lipsum}
\usepackage{xcolor}
\usepackage[normalem]{ulem}

\newcommand{\veloce}{{\texttt{VELOCE}}}
 
\newcommand{\ms}{\,m\,s$^{-1}$} 
\newcommand{\cmed}{{$\widetilde{C}$}}

\defcitealias{veloce}{A24}
\defcitealias{Netzel2024}{N24}
\defcitealias{Netzel2024b}{N25} 
\defcitealias{Binnenfeld2022}{B22}

\begin{document}
\makeatletter
\ifdefined\linenumbers
  \linenumbers
  \def\makeLineNumber{\relax} 
\fi
\makeatother

   \title{The VELOCE modulation zoo III.}

   \subtitle{Detecting additional pulsation modes in optical spectra of classical Cepheids using semi-partial distance correlation periodograms}

   \author{
          K. Barbey\inst{1}, R.I. Anderson\inst{1}, G. Viviani\inst{1},  H. Netzel\inst{2}, A. Binnenfeld\inst{3}, S. Zucker\inst{3, 6}, S. Shahaf\inst{4}, X. Dumusque\inst{5}
          }

   \institute{Institute of Physics, \'Ecole Polytechnique Fédérale de Lausanne (EPFL),
              Observatoire de Sauverny, 1290 Versoix, Switzerland\\
              \email{kentbarbey2598@gmail.com}, \email{richard.anderson@epfl.ch}
        \and
            Nicolaus Copernicus Astronomical Center, Polish Academy of Sciences, Bartycka 18, 00–716 Warsaw, Poland
         \and
             Porter School of the Environment and Earth Sciences, Raymond and Beverly Sackler Faculty of Exact Sciences, Tel Aviv University, Tel Aviv 6997801, Israel 
        \and
            Department of Particle Physics and Astrophysics, Weizmann Institute of Science, Rehovot 7610001, Israel 
        \and
            Astronomy Department of the University of Geneva, 51 Ch. des Maillettes, 1290 Versoix, Switzerland
        \and
            School of Physics and Astronomy, Raymond and Beverly Sackler Faculty of Exact Sciences, Tel Aviv University, Tel Aviv 6997801, Israel\\
             }

   \date{Received: 11.12.2024 / Accepted: 14.04.2025}
 
  \abstract
  {
  Known for their large amplitude radial pulsations, classical Cepheids are critical standard candles in astrophysics. However, they also exhibit various pulsational irregularities and additional signals that provide deeper insights into their structure and evolution. These signals appear in spectroscopic observations as shape deformations of the spectral lines. Using semi-partial distance correlation periodograms, we analysed high-precision spectroscopic data from the \texttt{VELOCE} project for four stars: $\delta$ Cephei ($\delta$ Cep), BG Crucis (BG Cru), X Sagittarii (X Sgr), and Polaris. For $\delta$ Cephei, our control star, only the main radial mode was detected, confirming its stability and suitability as a benchmark for the method. In BG Crucis, a strong additional signal at $\sim 3.01$ days was identified, likely linked to line splitting. X Sagittarii exhibited dominant additional signals, notably one at $\sim 12.31$ days, also associated with significant line splitting. Polaris revealed multiple low-frequency signals, with the most prominent candidate at $\sim 59.86$ days, which might be linked to the star's rotation period. We explored the semi-partial distance correlation periodograms by incorporating cross-correlation functions (CCFs) and their variants, such as the median-subtracted CCFs, which improved the sensitivity to variations in line shape. In particular, the latter enabled the faithful detection of primary and additional signals present in the 1D spectra of fainter stars and low-amplitude pulsators. The semi-partial distance correlation periodograms demonstrated their utility for isolating signals associated with line shape variations; although, the analyses were complicated by the presence of artefact subharmonics and a visible low-frequency power increase for Polaris and BG Crucis. This study underscores the method's potential for finding new and unexpected signals as well as detailed analyses of Cepheid pulsations and opens new pathways for asteroseismic investigations.}

   \keywords{Stars: oscillations (including pulsations) --
                Stars: variables: Cepheids --
                Methods: statistical --
                Techniques: spectroscopic
               }
    \titlerunning{The VELOCE modulation zoo III}
    \authorrunning{K. Barbey et al.}
    \maketitle

\section{Introduction}
    \label{intro}
    Classical Cepheids are well known for their large amplitude radial pulsations that render them useful as standard candles. However, Cepheids also harbour a host of interesting pulsational irregularities and additional signals of varying intensity and difficulty to detect that may provide further insights into their structure and evolution. The most well-known pulsational instability in Cepheids concerns nearly linear variations of the pulsation period $P$ over time (in s/yr) when observed over timescales lasting several decades \citep[e.g.][]{Szabados1983,Turner2006,Csornyei2022,veloce} that are typically attributed to secular evolution and generally agree with predictions from stellar evolution \citep{Anderson2016rot}. In addition, irregular and quasi-periodic short-term modulations of $P$ have also been reported, the former notably in overtone pulsators \citep[e.g.][]{Pietrukowicz2001,Evans2015,Suveges2018b,Suveges2018a} and the latter in long-period Cepheids, such as RS Puppis \citep[see also \citealt{veloce}]{Kervella2017}. Furthermore, we can also mention Blazhko-like periodic or quasi-periodic modulation in Cepheids, a seemingly rare occurrence, whose cause is still a mystery \citep{Molnar2013}. The frequency-modulated nature of these large amplitude primary pulsation modes hinders the detectability of lower-amplitude additional pulsation modes \citep{Suveges2018b}, which could cast open an asteroseismic window for unravelling the physics of Cepheids. 
    
    Most previous work to detect additional modes has employed precise space-based \citep{Plachy2021} and long (temporal) baseline ground-based photometric observations in search of these signals \citep[e.g.][and references therein]{rathour2021, smolec2023}. In many of the first-overtone Cepheids, the additional, low-amplitude signals forming period ratios of around 0.61--0.65 with the first overtone were detected, the so-called 0.61 signals. Interestingly, analogous signals forming the same characteristic period ratio were also found in the first-overtone RR Lyrae stars \citep[e.g.][]{benko2023, netzel_census, netzel_k2}. According to \citet{2016CoKon.105...23D}, the $0.61$ signals correspond to the harmonics of non-radial modes: the period ratio of $0.61$ is formed by $P_{1\rm O}/(0.5P_{\rm nr})$, where $P_{1\rm O}$ is the first-overtone period and $P_{\rm nr}$ is a non-radial mode period. That is to say, the subharmonics of the $0.61$ signals are the non-radial modes. Or, conversely, the signals observed are the harmonics of the non-radial mode. In the case of classical Cepheids, the non-radial modes have degrees 7, 8, or 9, while in the case of RR Lyrae stars, the degrees are 8 or 9. Another interesting group with additional signals includes the so-called 0.68 stars, in which the signal has a period longer than the first overtone period and both form a ratio of around 0.68 \citep{poretti2014, Suveges2018b}. Again, the analogous group is known among RR Lyrae stars \citep[e.g.][]{netzel2015b, netzel_k2}. The origin of $0.68$ signals is not as thoroughly understood at this time, as the suggested interpretations encounter challenges \citep[see discussion in ][]{Netzel2015a,2016CoKon.105...23D,benko.kovacs2023}. 

    Additional periodic or quasi-periodic phenomena also manifest in spectroscopic observations and radial velocity (RV) time series. In particular, \citet{Anderson2014rv,Anderson2020} show that overtone Cepheids exhibit long-term (decade or longer) modulations of their RV curves, whereas long-period Cepheids, such as $\ell$~Car and RS Puppis, exhibit fast cycle-to-cycle differences in RV curve shape and amplitude \citep[cf. also][]{Anderson2016c2c,Anderson2016vlti}. The VELOcities of CEpheids (\veloce) project \citep[][hereafter \citetalias{veloce}]{veloce} has now revealed these modulations in 31 Milky Way Cepheids, showing that they are a common feature of Cepheids whose detectability is mainly a question of precision and observational detail. In particular, the detection of non-radial modes requires sufficient phase sampling, number of observations ($N_{obs}$) and signal-to-noise ratio \citep[S/N; e.g.][]{Netzel2021}. 
    
    The \veloce\ modulation zoo makes use of high-precision RV measurements and spectroscopic observations from the \veloce\ project to investigate the detectability of additional signals in Cepheids and to identify promising strategies for systematic analyses. \citet[][hereafter \citetalias{Netzel2024}]{Netzel2024} report the detection of likely non-radial pulsation modes based on the analysis of numerical parameters that describe the asymmetry --bisector inverse span (BIS)--, full width at half maximum (FWHM), and depth (contrast) in four overtone Cepheids. Three of which belong to the $0.61$ group (BG~Cru, QZ~Normae, and V391~Normae), whereas V0411~Lacertae belongs to the $0.68$ group. In a follow-up study, \citet[][hereafter \citetalias{Netzel2024b}]{Netzel2024b} consider the relation between such signals and line splitting features observed in X~Sgr, BG~Cru, and other Cepheids \citep[cf. also][]{Kovtyukh2003}. \citetalias{veloce} and \citetalias{Netzel2024} thus demonstrate the general detectability of additional signals detected using {\it Kepler} and TESS \citep{Plachy2021} using \veloce\ observations. Additionally, \citetalias{veloce} and \citetalias{Netzel2024} confirmed that the modulated variability of Cepheids primarily changes line shapes rather than line positions (i.e. Doppler shift) \citep[see e.g. ][]{Anderson2016c2c,Anderson2019,Anderson2020}. Hence, observed RV modulations originate from line shape variations that do not occur in phase with the dominant radial mode, and the characterisation of such signals should focus on line shapes, rather than positions.

    Semi-partial distance correlation periodograms have been recently developed to separate spectral line displacements due to Doppler shifts from line shape variations due to stellar activity or pulsations \citep[][hereafter \citetalias{Binnenfeld2022}]{Binnenfeld2022}. In a first test based on optical spectra of two Cepheids, Binnenfeld et al. show that the corresponding shape and shift periodograms successfully distinguished periodicities associated with the main pulsation mode and an orbital signal in two Cepheids, $\beta$~Doradus and S~Muscae. Interestingly, these periodograms can be computed directly using observed spectra and do not require the information present in line shapes to be reduced to just a few parameters, such as BIS, FWHM, and contrast. 
    
    This article aims to exploit the semi-partial distance correlation periodograms sensitive to shape variations (henceforth: shape periodograms) presented by \citetalias{Binnenfeld2022} to probe the spectroscopic detectability of additional, possibly non-radial, pulsation modes in Cepheids. To this end, we analysed four very bright stars with a large number of high S/N observations from \veloce: $\delta$~Cep, BG~Cru, X~Sgr and Polaris. In doing so, we seek to lay the groundwork for further in-depth study of such additional signals using larger telescopes (for better S/Ns) and instruments that feature higher line shape stability and resolution. 
    
    The article is structured as follows. Section\,\ref{sec:data} first describes the data collected as well as data products derived from the observations (Sect.\,\ref{sec:obs}) and in turn summarises the methodology of the shape periodograms (Sect.\,\ref{sec:method}). Section\,\ref{sec:results} presents our results for the four stars individually (Sect.\,\ref{sec:stars}) and in turn considers the results with respect to one another (Sect.\,\ref{sec:overview}). Section\,\ref{sec:discussion} discusses the performance of the shape periodograms (Sect.\,\ref{sec:performance}) as well as the possible nature of the signals found (Sect.\,\ref{sec:signals}). The final Sect.\,\ref{sec:summary} summarises the results and presents our conclusions.

\section{Data and methods}\label{sec:data}

    This section is divided into  two parts. Sect. \ref{observations} gives a description of the observations, the star sample, and their relevant parameters and Sect. \ref{sparta} describes the functioning and application of the public python package SPectral vARiabiliTy Analysis (\texttt{SPARTA}), used throughout this work, to our data set.
    
    \subsection{Observational data used\label{sec:obs}}
    \label{observations}
    
\begin{figure*}[hbtp]
        \centering
        \includegraphics[width=0.49\hsize]{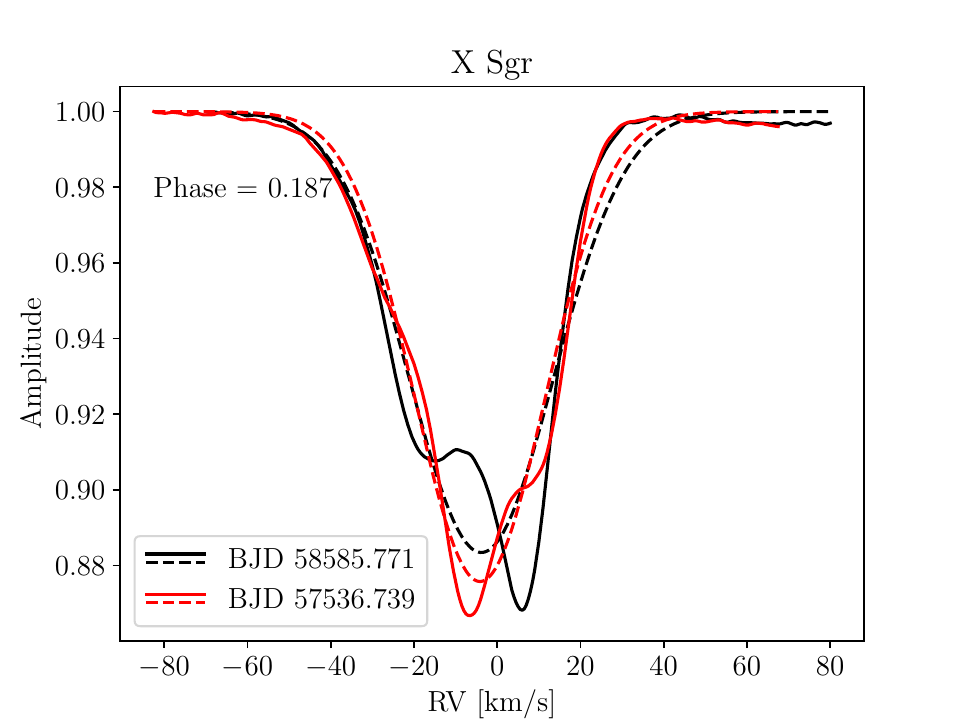}
        \includegraphics[width=0.49\hsize]{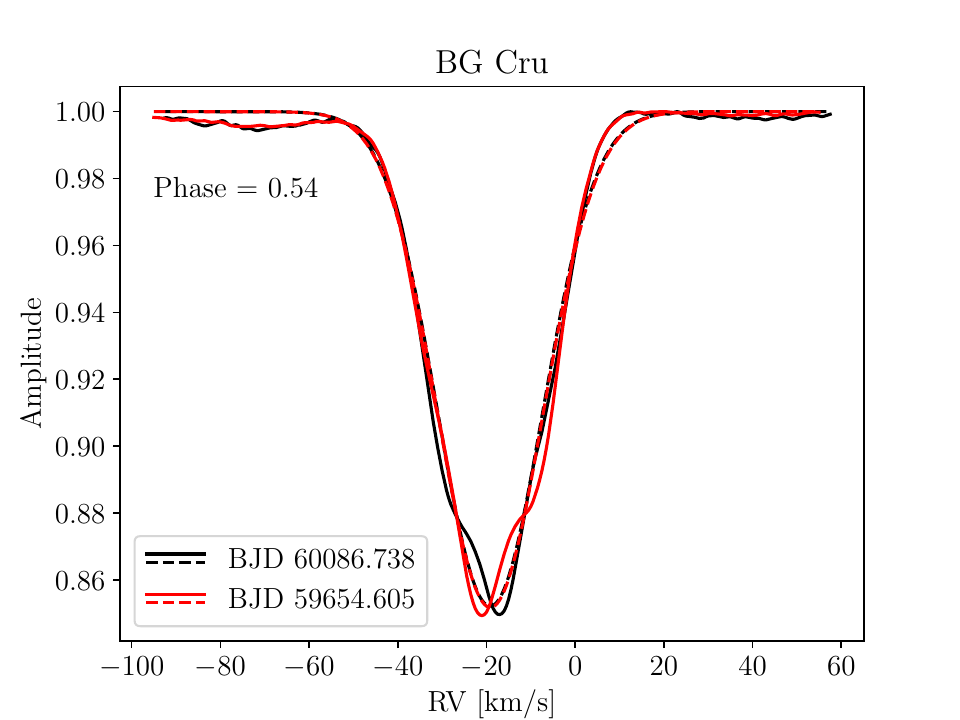}
        \caption{CCF profiles (solid lines) of X Sgr (left) and BG Cru (right) observed at the same phase during different epochs and their respective Gaussian fit (dashed lines). Cycle-dependent line splitting patterns cause systematic error for RVs derived via fitting Gaussians \citep{anderson}. This phenomenon is much stronger in X Sgr where the systematic error between the two epochs is clearly more discernible than in BG Cru.}
        \label{fig:line_splitting}
    \end{figure*}

    We analyse optical high-resolution spectroscopic observations of four Cepheids collected as part of the \veloce\ project \citepalias{veloce}. Specifically, we selected four bright naked-eye Cepheids, for which high S/N spectra were available in abundance. $\delta$~Cephei was chosen as a control star, for which we do not expect to identify additional signals beyond the fundamental radial pulsation mode, since fundamental mode Cepheids of relatively short pulsation period ($5-7$ d) are known to be particularly stable pulsators, that is, they do not typically present significant changes in period or RV curve shape. Three additional stars, X Sgr, BG Cru, and Polaris were chosen because they are known to exhibit interesting additional phenomena of different types and timescales. X Sgr is strongly affected by line splitting \citep{Mathias2006,anderson} and exhibits an additional 12 d periodicity in CCF parameters BIS and FWHM \citepalias{Netzel2024b}. BG Cru exhibits weak line splitting and a 0.61 mode \citep{anderson,Usenko2014} with periods close to the main radial mode. Figure\,\ref{fig:line_splitting} illustrates line splitting for X~Sgr and BG~Cru for two observations taken at virtually identical phase (cf. \citetalias{Netzel2024b} for further details). Polaris is a well-known single-lined binary ($P\sim 30 $yr) and is the oddball among Cepheids showing multi-periodic non-radial pulsation at large periods ($P\geq 20$ d) and has been the cynosure of several studies in the past \citep{Dinshaw1989,Hatzes2000,Lee2008,Anderson2019,Torres2023}.

    \begin{table*}
        \centering
        \caption{Summary of the data set used. \label{param}}
        \resizebox{\textwidth}{!}{
        \begin{tabular}{llllllll}
        \toprule
        Star            &RA (J2000) [h:m:s]&DEC (J2000) [d:am:as]&G [mag]&$P_{\rm r}$ [d]& Instrument & Baseline [BJD-2.4M]  & $N_{\mathrm{obs}}$    \\ 
        \midrule
        $\delta$ Cep &22:29:10&+58:24:54&3.8&5.366267 & HERMES     & $55816 - 58554$           & 263     \\
        BG Cru   &12:31:40&-59:25:26&5.3&3.342540      & C14         & $57155 - 60536$  & 508       \\
        X Sgr  &17:47:33&-27:49:50&4.3&7.012805         & C14      & $57153 - 60158$   & 185       \\
        Polaris &02:31:49&+89:15:50&-&3.972013    & HERMES     & $58239 - 60382$   & 290   \\ 
        \bottomrule
        \end{tabular}}
        \tablefoot{CCF profiles data for the four stars. The consecutive columns provide the star name, the coordinates, the intensity-averaged \textit{Gaia} DR3 G-band magnitude, the \veloce \, period, the instrument, the BJD range, and the number of observations. S/N averages for CCFs were computed using a 2$\sigma$-clipping and 10 iterations. Average S/N for 1D spectra are estimated near the $60$th order for Coralie and HERMES.}
    \end{table*}

    \begin{figure*}
        \centering
        \hspace{-1.3cm}
           \includegraphics[width=17.5cm]{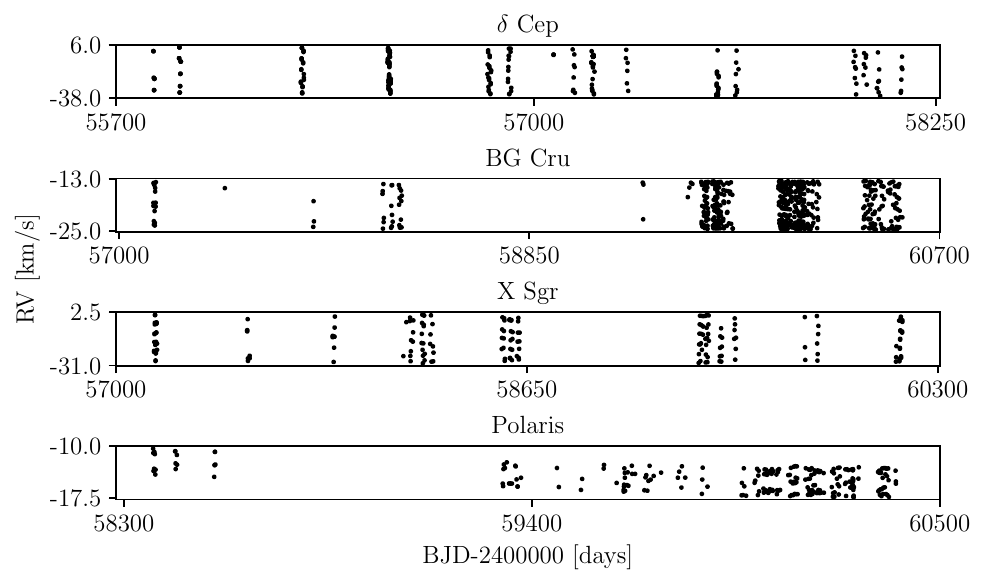}
       \caption{Sampling of spectroscopic observations for the four Cepheids. The x-axis represents the Barycentric Julian Date (BJD)-2400000 and the y-axis shows the measured radial velocities. We note that Polaris' time series shows a clear downward trend due to orbital variation ($P\sim 30$ y).}
       \label{sampling}
    \end{figure*}
        
    Northern hemisphere targets were observed with the HERMES high-resolution échelle spectrograph mounted on the 1.2m Mercator telescope at the Roque de los Muchachos observatory on the island of La Palma, Spain \citep{Raskin2011}. To achieve the best spectral resolution and throughput, all observations were made using the high-resolution fibre (HRF mode, $R\sim 85'000$). Since HERMES underwent modifications to its optical system in April 2018, we consider HERMES observations before and after this date as though they had been obtained using different instruments, that is, we analyse the data separately.
    Southern hemisphere targets were observed with the high-resolution ($R\sim 60'000$) Coralie \'echelle spectrograph mounted on the Euler Swiss telescope at La Silla Observatory in Chile. Since Coralie's optical system was modified in November 2014, observations before and after this date are analysed separately. The instrument is referred to as C07 (between 2007 and 2014) or C14 (after 2014) respectively. The absolute wavelength calibration was provided by ThAr lamps. Simultaneous drift corrections were available on Coralie from intra-order ThAr or using a Fabry-P\'erot \'etalon spectra. On both HERMES and Coralie, care was taken to avoid large ($> 100$\,\ms) drifts of the wavelength solution by recalibrating the instrument during the night.
    
    Wavelength-merged spectra were prepared by the dedicated instrument pipelines following standard recipes. We restricted our analysis of wavelength-merged spectra to the wavelength range $5000-5800$ \AA, seeking to maximise wavelength coverage to enhance any signals, while remaining in a region that is insensitive to tellurics and interstellar medium lines (sodium doublet). We avoid the bluer regions of the spectral range for two reasons. First, the normalisation procedure performs poorly due to the high density of lines. Second, the efficiency of the instruments is a strong function of the wavelength and plateaus at $\sim 5000 $  \AA\, \citep[see e.g. ][]{Raskin2011}.
    
    We normalised the spectra using the pipeline embedded in \texttt{SPARTA} (see Sect. \ref{discussion} and Appendix \ref{appendixB} for more details). Specifically, we resampled the spectra on a linear scale so that the wavelength values are evenly spaced. Then, we apply a Butterworth band-pass filter in order to remove the low-pass instrumental response and the high-pass instrumental noise. Finally a Tukey window is applied to smooth the data and reduce spectral leakage in view to improve the spectral analysis.
    
    In addition to the merged spectra, we analyse cross-correlation functions \citep[CCFs]{1996A&AS..119..373B,Pepe2002} computed using a G2 spectral line mask. CCFs  have several advantages for the detectability of additional signals over wavelength-merged spectra (K. Barbey, Master thesis, EPFL, 2024), such as enhanced S/N and the elimination of confounding spectral features, including telluric lines and the shape of the observed spectral envelope. A downside, however, could be the weighting of spectral lines introduced by the G2 line mask. For this reason, we decided to analyse both wavelength-merged spectra and CCFs and to compare the outputs. RVs were determined by fitting Gaussian profiles to the CCFs as done in \citetalias{veloce}. The measured RVs are defined as the centre positions of Gaussian profiles fitted to the resulting high S/N CCF profiles. Since \veloce\ observations have continued beyond the cut-off date for \veloce's first data release (March 2022), we included here new observations obtained since then. The observations are summarised in Table \ref{param} and the sampling is presented on Fig. \ref{sampling}.
    
    In the following, we write $C(v)$ for CCF profiles where $v$ is the radial velocity. In a time series of $N$ observations, the i-th fitted Gaussian is then defined as follows \citep{Pepe2002}:
    \begin{equation}
        g_i(v) = 1-A\exp{\left(-\dfrac{(v-\bar{v})^2}{2\sigma^2}\right)},
    \end{equation}
    where $A$ denotes contrast and $\sigma$ is related to the FWHM via: $\text{FWHM} = 2\sqrt{2\ln{2}}\cdot\sigma$.
    
\begin{figure*}
        \centering
           \includegraphics[width=17cm]{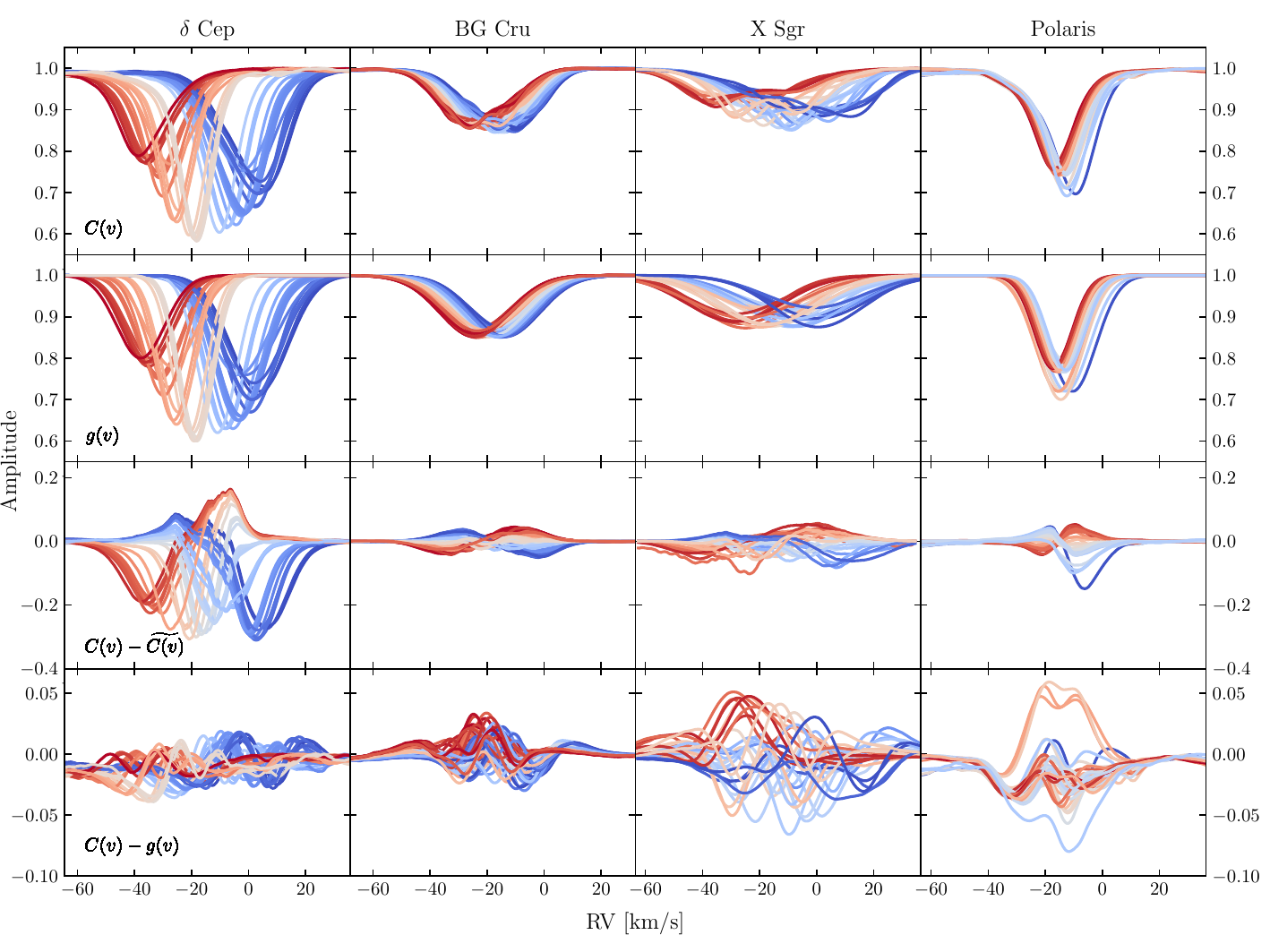}
       \caption{Subset examples of normalised CCF profiles, their Gaussian fit, and residual CCF profiles computed by subtracting the median CCF or their respective Gaussian fit to each one of them. The colour coding traces the pulsation phase. Shape modulations are computed with the curves `at rest', centred around the mean RV. For readability, the curves are shown in their original non-centred frame. X Sgr and $\delta$ Cep are fundamental mode pulsators whereas BG Cru and Polaris are first overtone ones. Line-splitting is noticeable in BG Cru's and X Sgr's $C(v)$ data sets.}
       \label{ccfs_examples}
    \end{figure*}

    Analogously to prewhitening procedures used in time series analysis, modifications of the input spectral shapes can enhance certain signals and diminish others. In addition to the original CCF profiles and the normalised spectra, we tested three variations on the CCF profiles as spectral shape inputs as illustrated in Figure \ref{ccfs_examples}:
    \begin{enumerate}
        \item the Gaussian fits, $g_i(v)$,
        \item median-residual CCFs obtained by subtracting the point-wise median of all CCF profiles, $C_i(v)-\widetilde{C(v)}$,
        \item Gaussian-residual CCFs obtained by subtracting the Gaussian fit from each CCF, $C_i(v)-g_i(v)$ .
    \end{enumerate}
    where $i$ represents the $i$-th element in the time series. By construction and contrary to the median and Gaussian-residual CCF profiles, Gaussian fits do not contain information on line asymmetry because they are symmetric profiles. 

    \subsection{Semi-partial distance correlation periodograms\label{sec:method}}
    \label{sparta}
    Pulsation phenomena in classical Cepheids, both radial and non-radial, manifest in their spectra as modulations in the shape of the spectral lines. Since CCF profiles are commonly interpreted as `average line profiles' pondered by their depth in the numerical mask \citep{2001A&A...379..279Q,anderson}, it is unsurprising that they, too, exhibit such line shape variations. CCFs are often analysed under the scope of so-called line shape indicators such as BIS, FWHM, and the contrast or depth (A) which are straightforward parameters to use in periodograms. They can identify and quantify CCF profiles shape periodicities \citep[e.g.][]{Anderson2016c2c,Anderson2019,Netzel2024}. However, they contain a reduced amount of information and correlation between themselves.
    
    The `shape' (i.e. line shape variation sensitive) and `shift' (i.e. line shift variation sensitive) periodograms introduced by \citetalias{Binnenfeld2022} present a more general approach to detecting periodic signals based on line shape deformations and line shifts. They rely on the detection of modulations in the line profiles themselves without resorting to the extraction of a single-valued parameter per observation.  The method, implemented in the public Python package \texttt{SPARTA}\footnote{\url{https://github.com/SPARTA-dev/SPARTA/tree/master}}, is here summarised. A detailed prescription on the periodogram calculation can be found in \citetalias{Binnenfeld2022} and references therein.
    
    Distance correlation is a generalised measure of statistical dependence between two random variables. Unlike Pearson's correlation, it is sensitive to non-linear dependencies and can be used between random variables of different dimensions. Similarly, semi-partial distance correlation is the generalised counterpart to semi-partial correlation. Stellar activity manifesting as spectral-shape variability can obscure orbital signals. Conversely, the binary nature of most Cepheids makes the detection of additional pulsations more difficult. By using semi-partial distance correlation between the phases, the spectral shapes, and the RVs, the influence of one variability over the other can be controlled. 
    
    Let us assume we have three time series of measured values $\mathbf{x}$, $\mathbf{y}$ and $\mathbf{z}$ each drawn from a joint distribution of random variables $X$, $Y$ and $Z$ that cannot be assumed to be independent. Then the Pearson correlation coefficient between $\mathbf{x}$ and $\mathbf{y}$ controlled from the influence of $\mathbf{z}$ on both of the first two variables is called the partial correlation. Practically, this means separately fitting two linear relationships: one between $\mathbf{x}$ and $\mathbf{z}$ and one between $\mathbf{y}$ and $\mathbf{z}$. The partial correlation coefficient then corresponds to the correlation coefficient computed between the residual of these fits. Conversely, semi-partial correlation considers the effect of $\mathbf{z}$ on only one of the other two variables, $\mathbf{x}$ or $\mathbf{y}$. The semi-partial distance correlation coefficients then play the same role as the Fourier power in a Fourier-based periodogram. A higher coefficient suggests a stronger signal at the corresponding frequency.

    To construct the shape periodograms used in the following, we compute the semi-partial distance correlation coefficients between the phases computed for a grid of test frequencies and the spectral shapes controlled from the influence of the RVs (`shape' power) by shifting the spectra to a common mean velocity. Conversely, the shift periodogram is constructed by computing the semi-partial distance correlation coefficients with the spectral shapes as a nuisance variable. Although both periodograms are computed in the process, we do not show the shift periodogram as its use is confined to the detection of orbital signals and we are only interested in detecting additional stellar-activity related pulsation modes.

    We compute for each of the four stars five different shape periodograms using the spectral inputs described above:
    \begin{enumerate}
        \item $C$ periodograms based on the CCF profiles,
        \item $g$ periodograms based on the Gaussian fits,
        \item $\widetilde{C}$ periodograms based on the median-residual CCF profiles,
        \item $C_g$ periodograms based on the Gaussian-residual CCF profiles,
        \item $S$ periodograms based on the normalised spectra.
    \end{enumerate}
    The five different shape periodograms are computed for each star with a frequency range spanning up to the Nyquist frequency defined as the median of separations between the observations. We note that $g$ periodograms are insensitive to line asymmetry by definition. \cmed\, periodograms are sensitive to variations from the median CCF profile. Finally, $C_g$ periodograms may be particularly sensitive to line asymmetry. 

    The significance levels of each periodogram are defined by false-alarm levels (FAL) corresponding to probabilities of 0.1, 0.01, and 0.001. The levels are calculated using the permutation procedure (bootstrap) described in Sect. 2.2 of \citet{2020A&A...642A.146B}. Peaks with false-alarm probabilities (FAP) $< 0.001$ and no other signal within ten times the periodogram's sampling are considered significant. The distinction between true frequencies and aliases is made following Sect. 7.2 of \citet{Vanderplas2018}. Following \citet{Gregory2001} and \citet{Ivezic2014}, the uncertainty on the peak's frequency $\sigma_f$ is defined by its width with a first-order dependence on the number of samples $N$ and their average S/N, $\Sigma$,
    \begin{equation}
        \sigma_f \approx f_{1/2}\sqrt{\dfrac{2}{N\Sigma^2}}x,
    \end{equation}
    where $f_{1/2}\approx 1/T$, with $T$ the baseline, is the half-width at half-maximum.

\section{Results}
\label{sec:results}
    This section is divided into two parts: Section \ref{sec:stars} describes the results individually for each star and Sect. \ref{sec:overview} presents an overview of the results. Depending on the context, we either discuss periods ($P$) or frequencies ($f$). For each star, a specific index is attributed to a given detected signal to facilitate its identification, e.g. $f_{\rm FU} = 1/P_{\rm FU}$ corresponds to fundamental mode (FU), and $f_{\rm X} = 1/P_{\rm X}$ to the dominant additional mode detected. We note that our sample does not contain any classical multi-mode (beat) Cepheids, so the nature of any additional modes is not a priori known. For harmonics and subharmonics, our terminology is the following: $n\cdot f$ is called the $n$-th harmonic and $f/n$ the $n$-th subharmonic for a given frequency $f$.

    \subsection{Individual stars\label{sec:stars}}

    We present the periodograms for the four stars in Figs. \ref{delcep_results}, \ref{bgcru_results}, \ref{xsgr_results}, and \ref{alfumi_results}. Tables \ref{results_delCep},  \ref{results_bgcru}, \ref{results_xsgr}, and \ref{results_alfUMi} list the detected main pulsation signal, the additional signals, their (sub)harmonics, and signals that are combinations of the main pulsation signal and an additional signal. Aliases of these signals with the window function are listed in Table \ref{period_results_b} of Appendix \ref{appendixA}.

    \subsubsection{\texorpdfstring{$\delta$}{d} Cephei}  
    \label{delCep}
\begin{figure*}[hbtp]
           \centering
               \includegraphics[width=17cm]{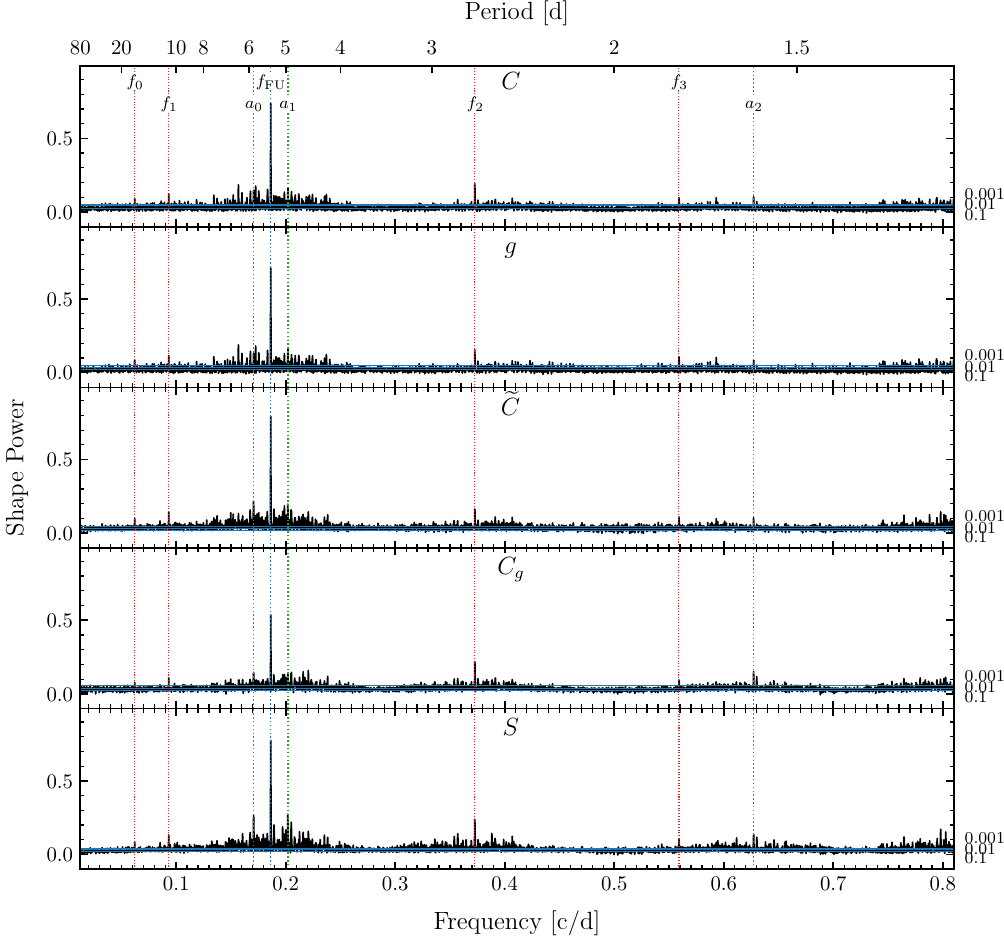}
           \caption{Shape periodograms for $\delta$ Cep. Rows correspond to: CCF profiles, their Gaussian fits, median-residual CCFs, Gaussian-residual CCFs, and normalised spectra. Significant peaks are marked with vertical lines: blue (radial pulsations and additional signals, $f_{\rm FU}$ for the fundamental mode, $f_{\rm 1O}$ for the first overtone, and $f_{\rm X}$ for others), red (harmonics, subharmonics, and combination signals; $f_i$, $i=0,1,2,\ldots$), and green (aliases; $a_i$, $i=0,1,2,\ldots$). Blue horizontal lines indicate FAL probabilities of $p=0.1$, $0.01$, and $0.001$ (bottom to top).}
           \label{delcep_results}
        \end{figure*}

        $\delta$ Cep is used as a control star for which we do not expect to detect additional frequencies beyond the main radial mode. For this star, we detect the fundamental radial pulsation frequency ($f_{\rm FU}$ in Table \ref{delcep_results}) as the most significant signal in all periodograms without exception. It is accompanied respectively by its second and third harmonics ($f_2$ and $f_3$) and subharmonics ($f_1$ and $f_0$). The second harmonic and subharmonic are significant in all periodograms, while the third ones stand out specifically in the $C$ and $g$ periodograms. Additionally, the third subharmonic is also significant in the $\widetilde{C}$ periodogram. The appearance of harmonics and, more importantly, subharmonics is discussed in Sect. \ref{sec:disc:subharmonics}. Three aliases are detected and listed in Table \ref{period_results_b}: $a_0$, $a_1$, and $a_2$. The first two are combinations of the fundamental mode frequency, $f_{\rm FU}$, with the second most prominent peak in the window power of $\delta$ Cep's time series at a period of 63.89 d. The third alias is a combination of the 1 d window signal with $f_2$.
        Evidentially, the line shape variability is sufficient to allow detection of the radial mode using all data sets and the shape periodograms. In particular, the $g$ data set also detects line shape variations because the parameters of the Gaussian fit react to significant changes in line asymmetry through the FWHM and the contrast. It is important to note that the line shape variability due to the radial mode is most pronounced in $\delta$ Cep compared to all other stars in the sample (Fig.\,\ref{ccfs_examples}).

    \subsubsection{BG Crucis}
    \label{BGCru}
        BG~Crucis is a first-overtone Cepheid that exhibits a small degree of line splitting \citep{Kovtyukh2003,anderson}. The known radial mode ($f_{\rm 1O}$) associated with BG~Cru's photometric and RV variability is detected in all periodograms but $C_g$ and it dominates in $g$. $f_{\rm 1O}$ is strongly suppressed in $C$ and yields the second strongest peak in $\widetilde{C}$ and $S$. The $S$ data set, contrary to CCF-based ones, shows an increase in low frequencies. This increase is aliased with the $1$ d window signal thus showing a cluster of peaks around $1$ d$^{-1}$ significantly above the threshold level. A second cluster, at $0.5$ d$^{-1}$, is also significant. We attribute it to the subharmonic of the $1$ d$^{-1}$ aliased low frequencies. The sharp power reductions at exactly $1$ d$^{-1}$ and $0.5$ d$^{-1}$ support this interpretation. The $g$ data set shows marginal signal split around $0.5$ d$^{-1}$ with the peaks being aliases of each other, similarly to \citetalias{Netzel2024} who finds a $2.03$ d signal based on FWHM and contrast. \citetalias{Netzel2024} attributes it to a non-radial mode of degree $l=7$ given the period ratio of $0.61$ with the main radial mode and according to \citet{2016CoKon.105...23D}. Similarly, we find a value of $2.03$ d for the most significant peak between the two. This signal is absent in the CCF-based periodograms. This suggests that the correlation between FWHM and contrast 
        might complicate the analysis when the full line profile is not analysed. Additionally, the presence of the low-frequency subharmonics around that frequency in the $S$ periodogram makes it difficult to determine whether the $2.03$ d signal detected in \citetalias{Netzel2024} is indeed present in that data set. 
                
        Interestingly, BG~Cru exhibits a very strong additional signal at a period of $P_{\rm X}\sim 3.01$\,d. This additional signal dominates over the main radial mode in the asymmetry-sensitive periodograms based on CCFs ($C$, $\widetilde{C}$, and $C_g$), whereas it is nearly of equal power as the radial mode in the S periodogram and fully suppressed in the $g$ periodogram. \citetalias{Netzel2024} recently identified $P_{\rm X}$ using Fourier analysis of time series of RV, FWHM, and BIS data, although the signal was not detected in the contrast. In \citetalias{Netzel2024b}, this signal was identified using a `hump' analysis of the split CCFs. In that approach, the hump formed between the two components of the line relative to the mean RV of the CCF is traced. The RV of the hump is then determined for each CCF profile and a time series of hump RVs is constructed and then analysed using standard time series analysis techniques. In the hump analysis of BG Cru's CCFs in \citetalias{Netzel2024b}, the 3 d signal appears as the highest-amplitude one. Consequently, it is assumed to be connected to the hump and thus to the line splitting. 
    
        \citet{moskalik2009b, moskalik2009} and subsequently \cite{kotysz2018} found evidence for non-radial oscillations in a significant fraction, approximately 9\%, of the first overtone Cepheids of the Large Magellanic Cloud present in the OGLE data sets. They observe that the secondary periodicities detected are located in 88\% of the cases close to the primary (radial) frequency and that the amplitude ratio $A_{\rm X}/{A_{\rm 1O}}$ never exceeds $0.1$. Moreover, in 84\% of these cases, the secondary peak has a lower frequency than the primary one. They argue that, given the frequency patterns observed and the period ratios induced, these secondary periodicities must be non-radial oscillations. Similarly, in our data sets of BG Cru, $f_{\rm X}$ is close to the main radial mode with $\Delta f \approx 0.033$. However, it appears right to it. Moreover, the amplitude ratios in the different data sets are ranged between $0.95$ and $7$. This difference can however be understood to be caused by the different methods used. The period ratio between $P_{\rm X} / P_{\rm 1O} \approx 0.90$ currently lacks a theoretical explanation in terms of a  non-radial mode. However, the fact that it is not in phase with $P_{\rm 1O}$ (see Fig. \ref{fig:line_splitting}) excludes a range of other physical origins such as shock or circumstellar environments \citep[cf.][]{Usenko2014}. Given the concordance of the periodicity between the hump analysis and the different periodograms, Fourier or distance correlation-based, the link between line splitting and this signal is very likely. \citet{Kovtyukh2003} tentatively attributed this line splitting (referred to as `bumps' and `asymmetries' in their text) to non-radial oscillations, potentially linked to resonances between radial modes in Cepheids pulsating with periods close to 3.2 d (see \citeauthor{Antonello1990} \citeyear{Antonello1990} for a description). Although the period of this additional signal is suspiciously close to 3 d, it is unlikely that it originates from a combination with the window function. The signal, which dominates four out of five periodograms, is also detected in the pre-2014 Coralie data set with different sampling, and is identified in \citetalias{Netzel2024b}'s hump analysis as well. 
    
        We note that harmonics of the main radial pulsation $f_{\rm 1O}$ are almost completely absent from the list of significant signals. The only detected harmonic, $f_6$, is barely identified in the $g$ periodogram and not in the others. On the other hand, several subharmonics of $f_{\rm 1O}$ and $f_{\rm X}$ are significant. We detect above the threshold level the second subharmonics of $f_{\rm 1O}$ and $f_{\rm X}$ ($f_3$ and $f_4$) and the third and fourth subharmonics of $f_{\rm X}$ ($f_2$, $f_1$). $f_0$ can be interpreted as either a tenth subharmonic or as a combination frequency of $f_{\rm X}-f_{\rm 1O}$. We find the former explanation unlikely although the absolute value of $f_0$ is closer to a tenth harmonic than $f_{\rm X}-f_{\rm 1O}$ is (an absolute difference of $6.75 \cdot 10^{-6}$ compared to $3.33 \cdot 10^{-2}$, respectively).
    
        We also detected signals ($f_5$ and $f_{7}$) that are combinations of the main radial mode, the additional frequency $f_{\rm X}$, and their (sub)harmonics. The presence of such combinations naturally appears in the frequency spectra if the S/N is high enough and the phase sampling dense \citep{Netzel2021}. The presence of unwanted subharmonics is discussed in Sect. \ref{sec:disc:subharmonics}.
    
    \subsubsection{X Sagittarii}
    \label{XSgr}
        X Sgr is a fundamental mode pulsator exhibiting strong line splitting \citep{Kovtyukh2003,Mathias2006,anderson}. Line shape periodograms detected the main radial mode at a period of $P_{\rm FU}\sim 7.01$ d using the full spectra ($S$), the Gaussian fits, and the $\widetilde{C}$ data sets. The full CCFs ($C$) do not yield a significant detection of this signal and there is no power in the CCFs once the Gaussian fit is subtracted ($C_g$). Hence, line shape variations due to the radial mode $f_{\rm FU}$ are subdominant compared to line shaping effects occurring at other frequencies. Subtracting the median CCF to construct the $\widetilde{C}$ data set enhances the shape variability stemming from the radial mode, which also significantly alters the depth and width of line profiles across the pulsation cycle. However, subtracting the Gaussian fits, which account for changes in line depth and width, from the CCFs ($C_g$) leaves the radial mode undetected in the shape periodograms.

        Apart from the main radial mode, we detected three significant signals at $f_{\rm X} = 0.081\, \mathrm{d^{-1}}$, $f_1 = 0.22\, \mathrm{d^{-1}}$, and $f_2 = 0.37\, \mathrm{d^{-1}}$. Notably, $f_1$ is the most dominant signal across all data sets but the $g$ one. However, based on the hump analysis of \citetalias{Netzel2024b}, which interprets $f_{\rm X}$ as the periodicity caused by the underlying phenomenon, we consider it an independent periodicity. Under this interpretation, $f_1$ and $f_2$ are combination frequencies, with $f_1$ satisfying $f_1 = f_{\rm FU} + f_{\rm X}$. Interestingly, $f_1$ exhibits a larger amplitude in the periodogram than its parent frequencies, one of which is a radial mode. This appears to contradict the findings of \citet{Balona2013} and \citet{benko.kovacs2023}, who, based on Fourier analysis, concluded that combination frequencies should not exceed their parent frequencies in amplitude when one of the parent is a radial mode. However, since shape periodograms are not Fourier-based, this constraint does not necessarily apply, leaving open the possibility that a combination frequency of a radial mode could surpass its parent in amplitude. Further investigation into the method itself is required to resolve this question. 
    
        Alternative interpretations remain plausible. In particular, \citetalias{Netzel2024b}'s `interpretation B' suggests that $f_2$ is the independent frequency. However, $f_2$ appears only mildly significant in the $\tilde{C}$ data set solely. Given these uncertainties, we remain agnostic about which frequency is the primary driver of the observed variability. Nevertheless, we attribute $f_{\rm X}$ to the noticeable line-splitting pattern, as it was identified through an analysis that traced the ‘hump’ between the main split components after subtracting an average line profile in \citetalias{Netzel2024b}. As is the case in BG Cru, we can exclude pulsation-induced shock waves as the origin of the line splitting pattern because it does not occur on the timescale of the main radial mode $P_{\rm FU}$, cf. Fig. \ref{fig:line_splitting} and Fig. 3 of \citetalias{Netzel2024b}. 
     
        $f_1$ and its combined nature with $f_{\rm FU}$ and $f_{\rm X}$ was previously detected by \citetalias{Netzel2024b} together with its subharmonic ($f_0$) and with the fundamental mode using standard Fourier analysis of the RV, BIS and FWHM curves of the same observations used here. However, $f_{\rm 1}$ (and $f_0)$ was also detected independently by \citet{smolec2018} using photometric observations from the BRITE satellites based on a total of 16 pulsation cycles ($\sim 112$ d) in 2017. 
        As noted by \citetalias{Netzel2024b}, the ratio of  $f_1 / f_{\rm FU}$ would place X~Sgr on a long-period extension of the so-called 0.61 Cepheids in a Petersen diagram (cf. top panel on Fig. 7 of \citet{Netzel2024b}). However, all previously reported 0.61 Cepheids are first overtone pulsators, whereas X Sgr's RV curve is typical of a fundamental mode pulsator (e.g. Fig. 8 of \citetalias{Netzel2024b}). Furthermore, if the hypothesis behind 0.61 Cepheids holds, that is they are the consequence of non-radial modes \citep{2016CoKon.105...23D}, then one would expect to find a different period ratio in the case of fundamental pulsators. Hence, despite the high amplitude of $f_1$ in the shape periodograms, and its previous photometric detection, we consider it unlikely for this frequency to represent a separate independent non-radial mode as described in \citet{2016CoKon.105...23D}.
   
    \subsubsection{Polaris}
        \label{alfUMi}
        Polaris is a first overtone, low-amplitude Cepheid pulsating with a period of $\sim$ 4 days, which is furthermore part of a triple system \citep{Evans2024}. As the nearest and brightest Cepheid, it has been intensely studied for a long time and offers many conundrums. Given its rapidly evolving photometric pulsation period, it is often considered a good candidate for a Cepheid on the first crossing of the instability strip, i.e. before core helium burning sets in \citep[e.g.][but it is important to note the updated mass in \citealt{Evans2024}]{anderson2018}. Polaris has also exhibited long-term amplitude variations in the past that seemed to have disappeared for several years in the 2010s \citep{ArellanoFerro1983,Dinshaw1989,Fernie1993,Bruntt2008,Anderson2019}.

        Of particular interest for this work are previously reported additional signals in the low-frequency regime ($P\gtrsim$ 20 d). Specifically, a period of $\sim 120$ d has been reported by \citet{Lee2008} and proposed as a rotation period. Other signals at $\sim 34.3$ d \citep{Kamper1998}, $\sim 40$ d \citep{Hatzes2000,Anderson2019},  $\sim45.3$ d \citep{Dinshaw1989}, and $\sim 60$ d \citep{Anderson2019} have been reported. \citet{Bruntt2008}, using observations from the star tracker on the WIRE satellite, concluded that additional periodicities reported by \citet{Dinshaw1989} and \citet{Hatzes2000} were likely spurious and argued for evidence of granulation on a timescale of $2-6$\,d based on a global power increase towards low frequencies.
        
        Here, our observations have the advantage of being part of a long-term, almost uninterrupted, data sequence that spans multiples of $120$ d. We detected the first overtone pulsation in the $\widetilde{C}$ and $S$ data sets. In the $C$, $g$ and $C_g$ data sets, the first overtone frequency is undetected. 
    
        The radial mode is detected in $\widetilde{C}$, but not from the CCFs themselves ($C$). Hence, any line shape variations associated with the radial mode are subdominant compared to other line shaping effects. However, the $\widetilde{C}$ data set does lead to a detection of the radial mode since the subtraction of the median line profile enhances the signature of line profile variations, notably variations in depth and width of the lines.
        
        We detected multiple additional signals in the low-frequency regime, from most to least significant, at $59.86$ d ($P_{\rm X2}$), $27.12$ d ($P_{\rm X6}$), $119.80$ d ($P_{\rm X1} = 2.00 \cdot P_{\rm X2}$), $37.86$ d ($P_{\rm X4}$), $33.48$ d ($P_{\rm X5}$), and $48.09$ d ($P_{\rm X3}$), all of which exceeded the $0.001$ false-alarm level in the periodograms of all data sets. The sole exceptions are $P_{\rm X1}$, which is absent in the $C_g$ data set, and $P_{\rm X3}$, absent in the $\tilde{C}$ one. We note that an increase in the baseline power of the periodogram towards longer periods ($\gtrsim 20$ d) renders the estimation of peak significance more complex. This low-frequency power increase is present in all stars in our data set, although it is most noticeable in Polaris and BG Cru, cf. Sect. \ref{sec:disc:subharmonics}. To more conservatively assess the significance of the peaks at $P \gtrsim 20$\,d in Polaris, we computed each peak's S/N relative to the immediate surroundings. The S/N was calculated using a box of 150 samples centred on the peak but excluding it. The box width and position were manually chosen to ensure a reliable noise estimate, and the S/N threshold for significant detection was set to $4.0$.
        
        Using this new criterion, we find that only one of the additional signals, $P_{\rm X2}$ (the highest amplitude peak), remains significant. Specifically, it passes the $4\sigma$ threshold for shape periodograms computed using the $g$ (S/N$=4.13$) and $\widetilde{C}$ (S/N$=4.75$) data set and remains marginally detected in the $C$ (S/N$=3.45$) and $S$ (S/N=$3.46$) data sets, whereas the $C_g$ data set yields a low S/N$=2.45$. We note that \citet{Anderson2019} had previously reported a similar frequency as dominant in the BIS parameter, although data gaps prevented a clearer identification at the time. The data set analysed here is not subject to such data gaps. The sampling presents a gap of two years between BJD 2458546 - 2459321 but otherwise was mostly uninterrupted in the BJD 2459321 - 2460381 period. Additionally, we note that \citet{Lee2008} had previously reported a signal at twice the period of $P_{\rm X2}$ (our $P_{\rm X6} = 119.8$\,d) based on RV measurements and considered this signal to be incoherent and possibly related to surface features, such as spots. However, the amplitude of the line shape periodograms suggest that the dominant period is indeed $P_{\rm X2}$. The recent detection of magnetic fields on Polaris \citep{Barron2022} and other Cepheids \citep{Barron2024} leads to the exciting possibility that a rotation period could be measured using magnetic maps based on spectropolarimetric observations, which may help to ascertain both the physical origin of the $\sim 60$\,d signal, or whether it is truly a $120$\,d signal.
    
        We did not detect the $\sim 40$ d signal reported by \citet{Hatzes2000} and \citet{Anderson2019} using the data set presented in Tab. \ref{param} although it is present in BIS (H. Netzel, private communication). It is however detected with this method in an independent smaller (161 observations) \veloce\, data set spanning BJD 2455817 - 2458443 which barely overlaps with the current one (22 observations). Moreover, several possible additional signals appeared below the S/N cutoff in the range of $27 - 48$\,d. This echoes reports of rather different similar periods that were generally considered non-coherent and thus potentially variable in time: $45.3$ d \citep{Dinshaw1989}, $34-45$ d \citep{Lee2008}, $34$ d \citep[using \citet{Kamper1998} data]{Hatzes2000}, $40.2$ d \citep{Hatzes2000}. However, the repeated appearance of the $40.2$\,d signal over the course of a few decades deserves attention and long-term monitoring, as does the persistence of the $60$ d signal.

    \subsection{Overview of results\label{sec:overview}}
    
        The previous subsection (\ref{sec:stars}) lists the results obtained per star. The sample of four stars exhibited quite different signals, on different timescales, and with different detections across the board. To assess the detectability of additional signals in spectroscopic time series observations of Cepheids using the (shape) semi-partial distance correlation periodograms, we seek here to provide an overview of our results, in particular by discussing the results as a function of which modes were detected, as well as differences between the data sets. Table \,\ref{tab:overview} provides an overview of our results for the four stars. It lists the primary radial frequency ($f_r)$, the frequency of the dominant additional signal ($f_{\rm X}$) detected, if any, and a short, pictographic overview of which data set yielded a significant and/or dominant detection.
        
        The photometrically dominant radial modes were detected using different data sets for different Cepheids, and for different reasons. In $\delta$ Cep, the detection of the radial mode based on all data sets was especially clear. We attribute this detection to the variation in equivalent width (EW) along the pulsation cycle. This would also explain why the  symmetric Gaussian fits allow the detection of $f_r$. Small period fluctuations could further contribute to this detection as they are common in Cepheids and have been noted in the paper that first reported the presence of a spectroscopic companion \citep{Anderson2015}. As expected, $\delta$ Cep did not exhibit any additional pulsation signals. We note that the shape periodograms are not sensitive to the known orbital period of $3451$\,d \citep{Shetye2024} by design.
    
        The radial mode was also detected in both line-splitting Cepheids in the sample, BG Cru and X Sgr, albeit using different data sets and likely because of the differing degrees of line splitting. Whereas the CCF profiles allowed for a weak detection of $f_r$ in BG Cru, the analogous signal did not exceed the threshold FAP in X Sgr, despite being visibly present in Fig.\,\ref{xsgr_results}.

        In both stars, $f_r$ was the dominant feature based on the Gaussian fits, as expected since the fits will vary in FWHM and depth to accommodate the changes in EW. The combination frequency $f_1 = f_{\rm FU} + f_{\rm X}$ dominated in the shape periodograms of the three remaining data sets of X Sgr. The radial modes were detected in $\widetilde{C}$ for BG Cru and X Sgr, as this data set enhances line shape variations by construction. However, other signals dominated in these data sets (see below). Conversely, $f_r$ was not detected in these two stars after subtracting the Gaussian fits ($C_g$), whereas it was detected using the full optical spectra.

        The radial mode was detected in Polaris based only on the $\widetilde{C}$ and $S$ data sets. We attribute this to the rather symmetric line profiles of this low-amplitude pulsator, which does not significantly vary its EW across the pulsation cycle. As a result, the $C$, $C_g$ and $g$ data sets yielded null detections for $f_r$. As expected from the analysis on the three other stars, subtraction of the median CCF in the $\widetilde{C}$ data set enhances the line shape variability due to the radial mode in Polaris. The full spectrum showed a very clean detection of $f_r$ as well as its main aliases and (sub)harmonics at $f_r/2$ and $2f_r$.

        Additional signals were detected by shape periodograms for BG Cru, X Sgr, and Polaris at timescales where they had been reported previously, albeit with some interesting differences among each other and with the literature. Importantly, the additional signals tend to dominate the shape periodograms based on CCFs when such signals are present.
    
        \begin{table*}[hbtp]
            \label{results_delCep2}
            \caption{Frequencies detected using shape periodograms for $\delta$ Cep.}
            \centering
            \begin{tabular}{@{}lllllllll@{}}
            \toprule
            \multirow{2}{*}{No.} &
              \multirow{2}{*}{Frequency [d$^{-1}$]} &
              \multirow{2}{*}{Period [d]} &
              \multicolumn{5}{c}{Amplitude} &
              \multirow{2}{*}{Identification} \\ \cmidrule(lr){4-8}
             &
               &
               &
              \multicolumn{1}{c}{$C$} &
              \multicolumn{1}{c}{$g$} &
              \multicolumn{1}{c}{$\widetilde{C}$} &
              \multicolumn{1}{c}{$C_g$} & 
              \multicolumn{1}{c}{$S$} &
               \\ \midrule
            $f_{\rm FU}$    & 0.1863267(2)  & 5.3669179  &\multicolumn{1}{c}{0.74}  &\multicolumn{1}{c}{0.71}&\multicolumn{1}{c}{0.79}  &\multicolumn{1}{c}{0.54}   &\multicolumn{1}{c}{0.77} &Fundamental Mode \\
            $f_0$    & 0.0621173(2) & 16.098569  &\multicolumn{1}{c}{0.094}&\multicolumn{1}{c}{0.087}&\multicolumn{1}{c}{-}&\multicolumn{1}{c}{-}&\multicolumn{1}{c}{-}& $f_{\rm FU}/3$     \\
            $f_1$    & 0.0932029(2)   & 10.729283  &\multicolumn{1}{c}{0.12}&\multicolumn{1}{c}{0.11}&\multicolumn{1}{c}{0.15}&\multicolumn{1}{c}{0.11}&\multicolumn{1}{c}{0.13}& $f_{\rm FU}/2$   \\
            $f_2$    &  0.3727071(2)  & 2.6830718   &\multicolumn{1}{c}{0.19}&\multicolumn{1}{c}{0.15}&\multicolumn{1}{c}{0.16}&\multicolumn{1}{c}{0.22}&\multicolumn{1}{c}{0.24}& $2f_{\rm FU}$    \\
            $f_3$    & 0.559088(2) &   1.7886284   &\multicolumn{1}{c}{0.097}&\multicolumn{1}{c}{0.11}&\multicolumn{1}{c}{0.11}&\multicolumn{1}{c}{-}&\multicolumn{1}{c}{-}& $3f_{\rm FU}$     \\
            \bottomrule
            \end{tabular}
            \tablefoot{Consecutive columns provide the frequency, the period conversion, and the amplitude of the peak for each periodogram in the same order as Fig. \ref{delcep_results} and the identification of the signal.}
            \label{results_delCep}
        \end{table*}

\clearpage

        \begin{figure*}[hbtp]
           \centering
               \includegraphics[width=17cm]{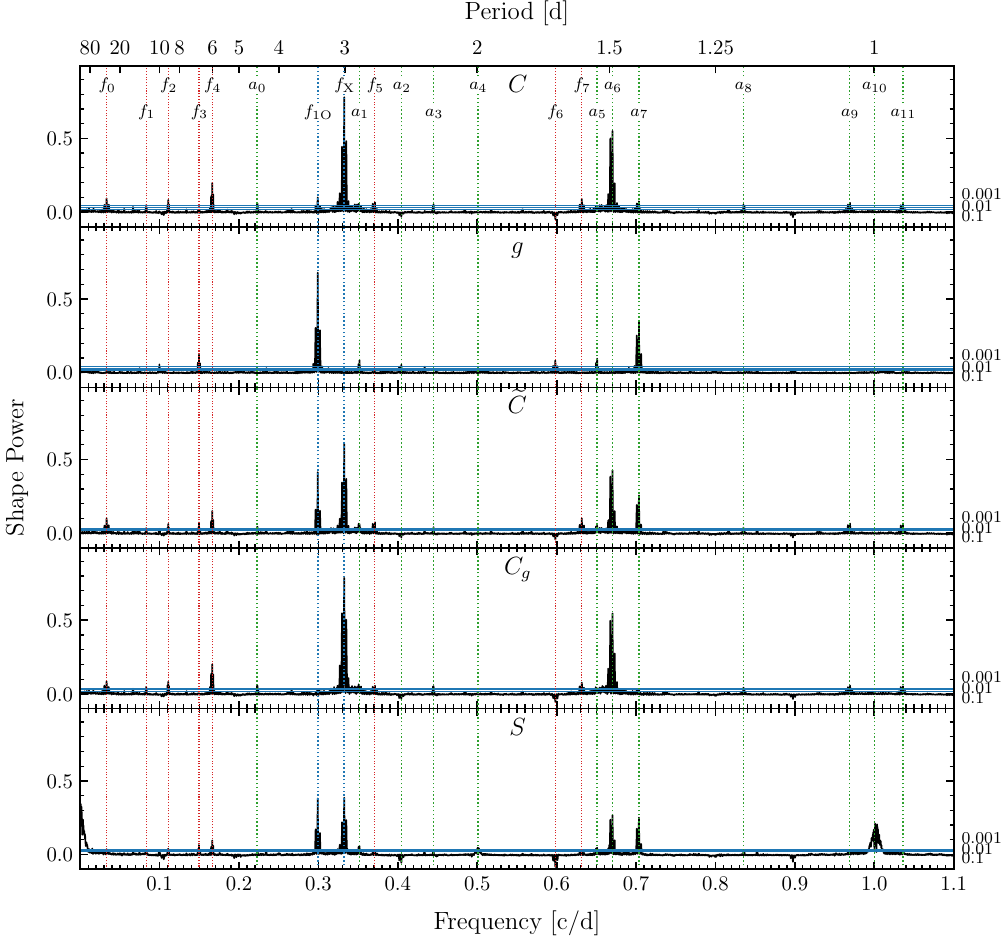}
           \caption{Same as Fig. \ref{delcep_results} but for BG Cru. The additional signal $f_{\rm X}$ is probably linked to line splitting.}
           \label{bgcru_results}
        \end{figure*}
    
        \begin{table*}[hbtp]
            \caption{Frequencies detected using shape periodograms for  BG Cru.}
            \centering
            \begin{tabular}{@{}lllllllll@{}}
            \toprule
            \multirow{2}{*}{No.} &
              \multirow{2}{*}{Frequency [d$^{-1}$]} &
              \multirow{2}{*}{Period [d]} &
              \multicolumn{5}{c}{Amplitude} &
              \multirow{2}{*}{Identification} \\ \cmidrule(lr){4-8}
             &
               &
               &
              \multicolumn{1}{c}{$C$} &
              \multicolumn{1}{c}{$g$} &
              \multicolumn{1}{c}{$\widetilde{C}$} &
              \multicolumn{1}{c}{$C_g$} & 
              \multicolumn{1}{c}{$S$} &
               \\ \midrule
            $f_{1\rm O}$ & 0.29917985(3) & 3.34247104  &\multicolumn{1}{c}{0.11}  &\multicolumn{1}{c}{0.67}&\multicolumn{1}{c}{0.43}  &\multicolumn{1}{c}{-}    &\multicolumn{1}{c}{0.38}  &First Overtone \\
            $f_{\rm X}$    & 0.33246066(3) & 3.00787470   &\multicolumn{1}{c}{0.77}  &\multicolumn{1}{c}{-}&\multicolumn{1}{c}{0.58}  &\multicolumn{1}{c}{0.77}    &\multicolumn{1}{c}{0.36}  &Additional Signal  \\ 
            $f_0$    & 0.03323932(3)  & 30.0848471  &\multicolumn{1}{c}{0.092}&\multicolumn{1}{c}{-}&\multicolumn{1}{c}{0.10}&\multicolumn{1}{c}{0.08}&\multicolumn{1}{c}{0.034}& $f_{\rm X}/10$ or $f_{\rm X}-f_{\rm 1O}$ \\
            $f_1$    & 0.08309935(3)  & 12.0337885  &\multicolumn{1}{c}{0.047}&\multicolumn{1}{c}{-}&\multicolumn{1}{c}{0.034}&\multicolumn{1}{c}{0.035}&\multicolumn{1}{c}{-}& $f_{\rm X}/4$                           \\
            $f_2$    & 0.11081296(3)  & 9.02421529   &\multicolumn{1}{c}{0.083}&\multicolumn{1}{c}{-}&\multicolumn{1}{c}{0.062}&\multicolumn{1}{c}{0.084}&\multicolumn{1}{c}{0.041}& $f_{\rm X}/3$                           \\
            $f_3$    & 0.14959978(3)  & 6.68450195   &\multicolumn{1}{c}{-}&\multicolumn{1}{c}{0.12}&\multicolumn{1}{c}{0.072}&\multicolumn{1}{c}{-}&\multicolumn{1}{c}{0.06}& $f_{\rm 1O}/2$                           \\
            $f_4$    & 0.16624018(3)  & 6.01539297   &\multicolumn{1}{c}{0.19}&\multicolumn{1}{c}{-}&\multicolumn{1}{c}{0.14}&\multicolumn{1}{c}{0.19}&\multicolumn{1}{c}{0.086}& $f_{\rm X}/2$   \\
            $f_5$    & 0.37112510(3)  & 2.69450907   &\multicolumn{1}{c}{0.066}&\multicolumn{1}{c}{-}&\multicolumn{1}{c}{0.075}&\multicolumn{1}{c}{0.055}&\multicolumn{1}{c}{-} &  $f_{\rm X} + f_0$  \\                        
            $f_6$    & 0.59834001(3)  & 1.67129054 &\multicolumn{1}{c}{-}&\multicolumn{1}{c}{0.081}&\multicolumn{1}{c}{-}&\multicolumn{1}{c}{-}&\multicolumn{1}{c}{-}& $2f_{\rm 1O}$ \\
            $f_{7}$    & 0.63162081(3)  & 1.58322838  &\multicolumn{1}{c}{0.087}&\multicolumn{1}{c}{-}&\multicolumn{1}{c}{0.097}&\multicolumn{1}{c}{0.077}&\multicolumn{1}{c}{0.028} & $f_{\rm 1O}+f_{\rm X}$   \\
            \bottomrule
            \end{tabular}
            \label{results_bgcru}
            \tablefoot{The line splitting pattern is attributed to $f_{\rm X}$. Subharmonics $f_0$ to $f_4$ are artefacts induced by the partial distance correlations (Sect.\,\ref{sec:disc:subharmonics}). }
        \end{table*}

\clearpage
    
        \begin{figure*}[hbtp]
           \centering
               \includegraphics[width=17cm]{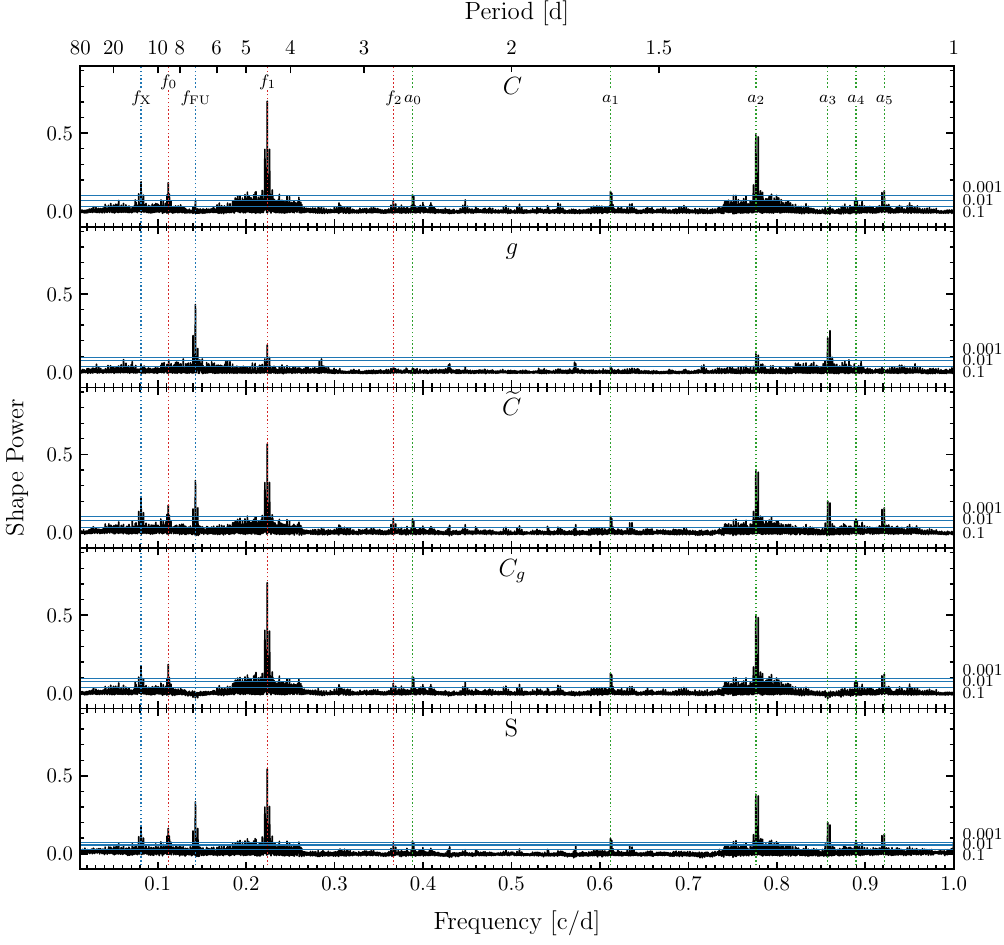}
           \caption{Same as Fig. \ref{delcep_results} but for X Sgr.}
           \label{xsgr_results}
        \end{figure*}
    
        \begin{table*}[hbtp]
            \caption{Frequencies detected using shape periodograms for  X Sgr.}
            \centering
            \begin{tabular}{@{}lllllllll@{}}
            \toprule
            \multirow{2}{*}{No.} &
              \multirow{2}{*}{Frequency [d$^{-1}$]} &
              \multirow{2}{*}{Period [d]} &
              \multicolumn{5}{c}{Amplitude} &
              \multirow{2}{*}{Identification} \\ \cmidrule(lr){4-8}
             &
               &
               &
              \multicolumn{1}{c}{$C$} &
              \multicolumn{1}{c}{$g$} &
              \multicolumn{1}{c}{$\widetilde{C}$} &
              \multicolumn{1}{c}{$C_g$} & 
              \multicolumn{1}{c}{$S$} &
               \\ \midrule
            
            $f_{\rm FU}$    & 0.14259569(4)   & 7.01283432  &\multicolumn{1}{c}{-}   &\multicolumn{1}{c}{0.43}&\multicolumn{1}{c}{0.33}  &\multicolumn{1}{c}{-}   &\multicolumn{1}{c}{0.33}  &Fundamental Mode \\
               $f_{\rm X}$    & 0.08120484(4)   & 12.3145361&\multicolumn{1}{c}{0.19}&\multicolumn{1}{c}{-}&\multicolumn{1}{c}{0.23}&\multicolumn{1}{c}{0.17}&\multicolumn{1}{c}{0.18}      & Additional signal                 \\
               $f_0$   & 0.11193918(4)   & 8.93342296 &\multicolumn{1}{c}{0.18}  &\multicolumn{1}{c}{-}&\multicolumn{1}{c}{0.18}  &\multicolumn{1}{c}{0.18}    &\multicolumn{1}{c}{0.16}  &$f_{\rm 1}/2$  \\ 
            $f_{\rm 1}$    & 0.22382770(4)   & 4.46772227  &\multicolumn{1}{c}{0.70}   &\multicolumn{1}{c}{0.17}&\multicolumn{1}{c}{0.57}  &\multicolumn{1}{c}{0.71}   &\multicolumn{1}{c}{0.54}  &$f_{\rm FU}+f_{\rm X}$\\
            $f_{\rm 2}$    & 0.36645056(4)   & 2.72888112  &\multicolumn{1}{c}{-}   &\multicolumn{1}{c}{-}&\multicolumn{1}{c}{0.090}  &\multicolumn{1}{c}{-}   &\multicolumn{1}{c}{-}  &2$f_{\rm FU} + f_{\rm X}$\\
            \bottomrule
            \end{tabular}
            \label{results_xsgr}
            \tablefoot{The line splitting pattern is attributed to $f_{\rm X}$.}
        \end{table*}

\clearpage

        \begin{figure*}[hbtp]
           \centering
               \includegraphics[width=17cm]{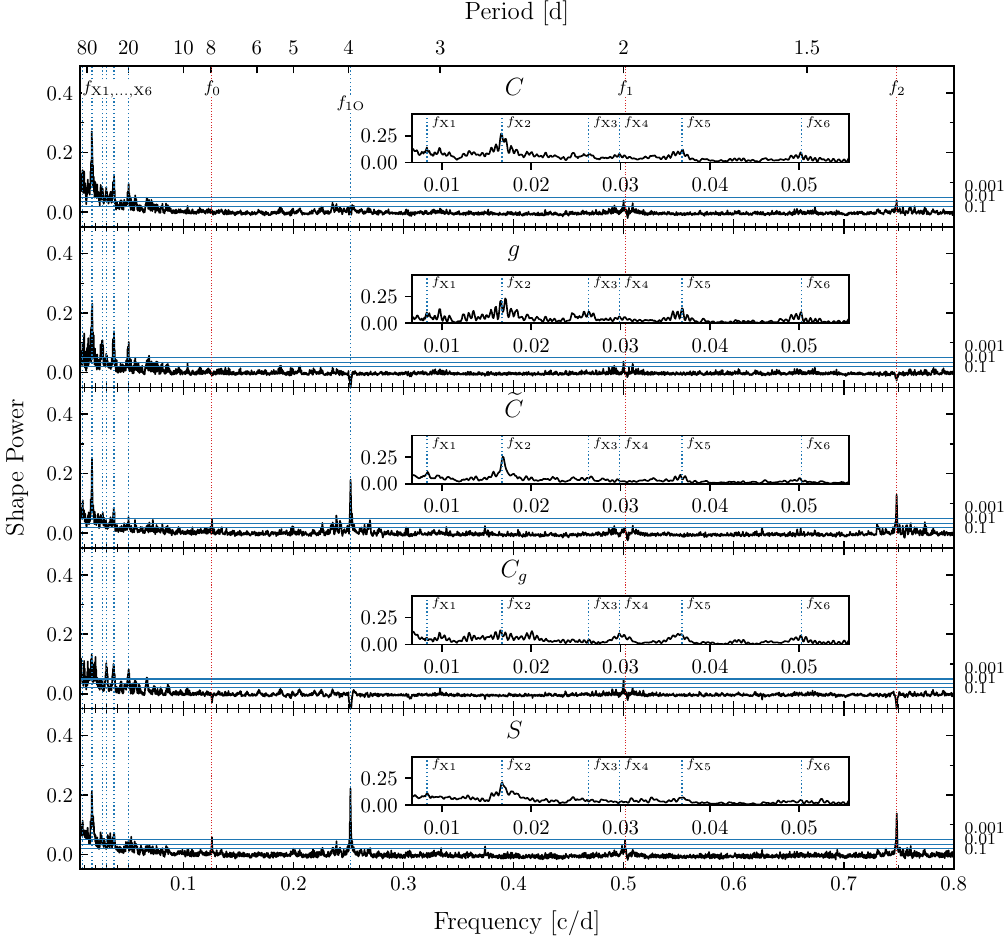}
           \caption{Same as Fig. \ref{delcep_results} but for Polaris. We note the increase of the power baseline for small frequencies.}
           \label{alfumi_results}
        \end{figure*}
    
        \begin{table*}[hbtp]
            \caption{Frequencies detected using shape periodograms for Polaris.}
            \centering
            \begin{tabular}{@{}llllllllll@{}}
            \toprule
            \multirow{2}{*}{No.} &
              \multirow{2}{*}{Frequency [d$^{-1}$]} &
              \multirow{2}{*}{Period [d]} &
              \multicolumn{5}{c}{Amplitude} &
              \multirow{2}{*}{Identification} \\ \cmidrule(lr){4-8}
             &
               &
               &
              \multicolumn{1}{c}{$C$} &
              \multicolumn{1}{c}{$g$} &
              \multicolumn{1}{c}{$\widetilde{C}$} &
              \multicolumn{1}{c}{$C_g$} & 
              \multicolumn{1}{c}{$S$} &
               \\ \midrule
            $f_{\rm 1O}$    &0.25175868(2) &3.97205765  &\multicolumn{1}{c}{-}  &\multicolumn{1}{c}{-}&\multicolumn{1}{c}{0.18}  &\multicolumn{1}{c}{-}    &\multicolumn{1}{c}{0.22}  &First Overtone \\
            $f_{\rm X1}$    &0.00834699(2)  &119.8036694    &\multicolumn{1}{c}{0.14}&\multicolumn{1}{c}{0.10}&\multicolumn{1}{c}{0.11}&\multicolumn{1}{c}{-}&\multicolumn{1}{c}{0.11}& $f_{\rm X2}$/2     \\
            $f_{\rm X2}$    &0.01670628(2)  &59.85774104    &\multicolumn{1}{c}{0.27}&\multicolumn{1}{c}{0.23}&\multicolumn{1}{c}{0.25}&\multicolumn{1}{c}{0.13}&\multicolumn{1}{c}{0.21}& Additional Signal       \\
            $f_{\rm X3}$    &0.02079552(2)  &48.08727186&\multicolumn{1}{c}{0.077}&\multicolumn{1}{c}{0.060}&\multicolumn{1}{c}{-}&\multicolumn{1}{c}{0.12}&\multicolumn{1}{c}{0.064}   &  Additional Signal $\dagger$     \\
            $f_{\rm X4}$    &0.02641568(2)  &37.85630784   &\multicolumn{1}{c}{0.067}&\multicolumn{1}{c}{0.11}&\multicolumn{1}{c}{0.069}&\multicolumn{1}{c}{0.046}&\multicolumn{1}{c}{0.063}& Additional Signal $\dagger$   \\
            $f_{\rm X5}$    &0.02986998(2)  &33.47842574    &\multicolumn{1}{c}{0.053}&\multicolumn{1}{c}{0.063}&\multicolumn{1}{c}{0.042}&\multicolumn{1}{c}{0.10}&\multicolumn{1}{c}{0.066}& Additional Signal $\dagger$  \\
            $f_{\rm X6}$    &0.03700585(2)  &27.12088438    &\multicolumn{1}{c}{0.12}&\multicolumn{1}{c}{0.14}&\multicolumn{1}{c}{0.084}&\multicolumn{1}{c}{0.10}&\multicolumn{1}{c}{0.072}& Additional Signal $\dagger$  \\
            $f_0$    &0.12587827(2)   &7.94418278     &\multicolumn{1}{c}{-}&\multicolumn{1}{c}{-}&\multicolumn{1}{c}{-}&\multicolumn{1}{c}{-}&\multicolumn{1}{c}{0.058}& $f_{\rm 1O}/2$         \\
            $f_1$    &0.50156918(2)  &1.99374292      &\multicolumn{1}{c}{-}&\multicolumn{1}{c}{-}&\multicolumn{1}{c}{-}&\multicolumn{1}{c}{-}&\multicolumn{1}{c}{0.049}& $2f_{\rm 1O}$         \\
            $f_2$    &0.74814635(2)  &1.33663687      &\multicolumn{1}{c}{-}&\multicolumn{1}{c}{-}&\multicolumn{1}{c}{0.13}&\multicolumn{1}{c}{-}&\multicolumn{1}{c}{0.14}& $3f_{\rm 1O}$         \\
            \bottomrule
            \end{tabular}
            \tablefoot{Polaris exhibits excess power at low frequencies ($P \gtrsim 20$\,d), which adds complexity to identifying additional signals. Candidate additional signals based on the global false alarm probability are labelled by a $\dagger$ symbol. $f_{\rm X1}$ is also significantly detected when taking into account the local envelope of the periodogram (see Sect.\,\ref{alfUMi}).}
            \label{results_alfUMi}
        \end{table*}

\clearpage

        \begin{table*}[hbtp]
            \centering
            \caption{Overview of mode detections per data set. \label{tab:overview}}
            \begin{tabular}{lllcccccccccc}
            \toprule
             & & & \multicolumn{2}{c}{$C$}  & \multicolumn{2}{c}{$g$} & \multicolumn{2}{c}{$\widetilde{C}$} & \multicolumn{2}{c}{$C_g$}  & \multicolumn{2}{c}{$S$} \\
            Star & $f_{\rm r}$ [d$^{-1}$] & $f_{\rm X}$ [d$^{-1}$] & $f_{\rm r}$? & $f_{\rm X}$? & $f_{\rm r}$? & $f_{\rm X}$? & $f_{\rm r}$? & $f_{\rm X}$? & $f_{\rm r}$? & $f_{\rm X}$? & $f_{\rm r}$? & $f_{\rm X}$? \\
            \midrule
            $\delta$ Cep & $0.19$ & $--$ & $\blacksquare$& $--$ & $\blacksquare$ & $--$ & $\blacksquare$ & $--$ & $\blacksquare$ & $--$  & $\blacksquare$ & $--$ \\
            BG Cru & $0.30$ & $0.33$ & $\square$ & $\blacksquare$ & $\blacksquare$ & $--$& $\square$ & $\blacksquare$ & $--$ & $\blacksquare$  & $\blacksquare$ & $\square$ \\
            
            X Sgr & $0.14$ & $0.08$ & $--$ & $\square$& $\blacksquare$ & $--$ & $\square$ & $\square$ & $--$ & $\square$  & $\square$ & $\square$ \\
            Polaris & $0.25$ & $0.017$ & $--$ & $\blacksquare$& $--$ & $\blacksquare$ & $\square$ & $\blacksquare$ & $--$ & $\blacksquare$  & $\blacksquare$ & $\square$  \\
            \bottomrule
            \end{tabular}
            \tablefoot{$f_{\rm r}$ denotes the main radial mode known from photometric variability, which is the fundamental mode for $\delta$~Cep and X~Sgr, and the first overtone for BG~Cru and Polaris. $\blacksquare$ denotes the dominant peak (in terms of significance), $\square$ denotes that that peak is present, but not dominant. $--$ indicates the absence of peak.}
        \end{table*}
        
        We note that the three significant additional frequencies ($f_{\rm X}$ for BG Cru and X Sgr; $f_{\rm X2}$ Polaris) reported are very different from one another, despite rather similar observational strategies employed for the three stars, which included nightly monitoring over at least two weeks at a time as well as entire seasons. Furthermore, the frequency ratios of $f_r / f_X$ are very different in each star and range from $< 1$ to $\gg 1$. Last, but not least, the additional signals were also detected using the full optical spectra ($S$). In BG Cru and Polaris, the amplitude of $f_{\rm X}$ was similar to $f_r$. In X Sgr, the largest amplitude was found for the combination frequency $f_1$, followed by $f_r$ and $f_{\rm X}$.
        
        To summarise the behaviour of the data sets, CCFs ($C$) are highly useful to detect additional signals based on spectroscopic observations and yield enhanced amplitudes at these interesting frequencies compared to the full optical spectra. This is likely a consequence of the increased S/N of CCFs compared to the restricted spectral range we've used, as well as a virtue of CCFs being free of several complicating factors, such as telluric or ISM lines, as well as time variable spectral envelopes. Fitted Gaussians ($g$) can detect signals that change the EW of the fit, and this was only the case in Polaris. $\widetilde{C}$ emphasises the line asymmetry and thereby enhances the radial mode signal. However, in the $\widetilde{C}$ data sets, additional signals generally still dominate the shape periodograms, and yield similar amplitudes as the original CCFs for the additional signals. The full optical spectra ($S$) allowed detection of both the radial mode and the additional signals with typically similar amplitude, although the radial mode was slightly stronger in BG Cru and Polaris and noticeably weaker or stronger in X Sgr compared to $f_1$ and $f_{\rm X}$, respectively.
        
    \section{Discussion\label{sec:discussion}}
        \label{discussion}

        This section is divided into two parts. In the first one, we assess the performance of these new periodograms on our data in comparison with past results that used different methods. Then, we discuss the nature of the additional signals detected.

        \subsection{Performance of the shape periodograms\label{sec:performance}}

        This subsection is divided into two parts. In Sect. \ref{sec:disc:comparison}, we compare our method and results with previous ones. In Sect. \ref{sec:disc:subharmonics}, we discuss more specifically the presence of sub(harmonics) and low-frequency power increase in some of our data sets.

        \subsubsection{Comparison with different methods}\label{sec:disc:comparison}

            We considered two main types of spectrosopic data: normalised 1D merged spectra and CCFs, as well as variations of CCFs. Concerning spectra, one of us (K. Barbey, EPFL Master thesis 2024) extensively investigated the impact of the selected wavelength regions and ranges, as well as spectrum normalisation techniques and other systematics of relevance. Notes on these issues are provided in Appendix \ref{appendixB}. CCFs were included in the analysis to benefit from enhanced S/N and because CCFs isolate the lines of interest from several factors present in real observed spectra that could negatively impact the periodicity search, including the shape of the continuum, telluric and ISM lines, contamination from the simultaneous ThAr or Fabry-Pérot wavelength reference, and lower S/N per resolution element. Nevertheless, normalised spectra nonetheless carry a great deal of information and may have some advantages over CCFs, such as the absence of the weighting that is introduced by the correlation template. Interestingly, we found that both CCF-based line profiles and the normalised spectra allowed to faithfully recover the additional signals. However, the amplitude of the additional signals was strongly boosted for CCF-based data sets compared to the full spectra, whereas the radial modes were usually strongly suppressed in the CCF-based data sets. We therefore expect that CCF-based data sets will be particularly suitable for fainter stars compared to these four naked-eye stars. In particular, the $\widetilde{C}$ data set appears to be promising, as it boosts the main signal by emphasising the line asymmetry variability due to the radial mode. Distinguishing the radial mode from other signals is usually straightforward when photometric light curves are available, such as from the ESA {\it Gaia} mission \citep[]{gaia_mission,gdr3,gdr3_vari,gdr3_cepheids}.

            A significant advantage of the present analysis over the analysis of individual CCF parameters, such as the depth, FWHM, and BIS \citep[cf. Fig.\,1 in][]{Anderson2016c2c}, is that it does not consider correlated quantities as independent variables. Nevertheless, an advantage of CCF parameters is that they can be calculated with good precision even from low S/N spectra. However, certain difficulties have arisen in their use, such as the detection of frequencies for only specific parameters (instead of all).
    
            Additional signals dominate in the shape periodograms. By comparison, a classical Fourier analysis of line shape parameters, such as BIS \citepalias[][]{Netzel2024b} identified these signals at lower S/N, and required pre-whitening by the radial mode for BG Cru. The separation into shape and shift periodograms in \texttt{SPARTA} performs an analogous operation for the radial modes, although they remain detectable in some cases. Indeed, the shift periodograms\footnote{See Data availability.} for our four stars and for all data sets show significant residual peaks at the main pulsation frequency, contrary to the periodograms presented for S~Mus and $\beta$~Dor in \citetalias{Binnenfeld2022}. These residual peaks were less pronounced in $S$ data sets of fundamental-mode pulsators (X Sgr, $\delta$ Cep and see K. Barbey, EPFL Master thesis 2024)) than in first-overtone Cepheids, which are known to present long-term amplitude and frequency modulations of their main pulsation \citep{Evans2014,Anderson2014rv,Anderson2018rvs,Suveges2018a}. Hence, the residual peaks may arise due to the instability of the radial pulsation period. Moreover, Gaussian-fitted RVs are biased with respect to the barycentric velocity of the star. We argue that these two reasons are probable causes of the presence of the residual peak. Indeed, the shape metric \citepalias[Eq. 14 in ][]{Binnenfeld2022} in the algorithm is calculated after the spectra are shifted according to their respective RV. Under the assumption that the Doppler shift and the line shape deformation are strongly associated in the case of a Cepheid's radial pulsation, the main pulsation frequency should not appear in the shift periodogram. Both causes could produce this residual signal.

            \subsubsection{On (sub)harmonics and the low-frequency power increase\label{sec:disc:subharmonics}}
            
                Figures~\ref{bgcru_results} (bottom panel) and \ref{alfumi_results} show a trend where the semi-partial distance correlation power increases at low frequencies. We attribute this trend to correlated noise, where consecutive measurements are more likely to deviate from the signal's expected value in the same direction. This correlated noise affects the periodogram through the folding process: at sufficiently low frequencies, the folding procedure aligns neighbouring points in neighbouring phases. Since the noise is correlated, these measurements are not statistically independent, resulting in an increased distance correlation score. This is the case for all periodograms of the phase distance correlation `family'. A similar effect was noted in previous work \citepalias[e.g.][]{Binnenfeld2022}. 

                Pulsation patterns that are not strictly sinusoidal produce harmonics in their periodograms. This is a well-known effect in many methods, including classical Fourier analysis. Similarly, phase distance correlation-based periodograms can exhibit significant peaks at both the harmonics and subharmonics of the detected signal. This behaviour can be understood in light of the phase-folding characteristics discussed in the previous paragraph. Statistical dependence between points in neighbouring phases of the true periodicity is likely to induce similar dependence in the corresponding phases of a harmonic or subharmonic of the true periodicity. This is conceptually similar to the patterns observed in the `box least squares' periodogram \citep{Kov02}. Subharmonic features tend to accompany the highest amplitude peak in the periodogram — whether that peak corresponds to the primary radial mode, an additional frequency, or a combination frequency. This behaviour is observed, for example, in BG Cru and Polaris, where our periodograms reveal fractional frequencies (e.g. near $0.5f$, and in some cases additional fractions). In contrast, true harmonics (i.e. integer multiples of the primary frequency) appear predominantly for the radial modes. Our method, by directly probing line shape variations, appears to accentuate these artefact subharmonics. Phase folding perspective further suggests that while the harmonics are phase-coherent and linked to the underlying radial mode, the subharmonics lack such coherence, which can indicate their non-physical origin in at least some cases. We acknowledge that when only a $0.5f$ signal is detected (as in some cases), alternative explanations cannot be entirely ruled out. We emphasise that further investigation is needed to definitively disentangle the effects involved. While our method's sensitivity to line shape variations is useful for such tasks, fully characterising the signals requires additional analysis and careful consideration of periodogram artefacts.

        \subsection{On the nature of the additional signals\label{sec:signals}}
            This subsection is divided into three parts. In the first one, we discuss the low-frequency increases in power in Polaris' data sets. In the second part, we review Polaris' long-period signals. Finally, in the third part, we discuss the signals apparently linked to line splitting in BG Cru and X Sgr.
            \subsubsection{Low-frequency increase in power}
                Several of our periodograms exhibited small power increases towards low frequencies, and these may be (partially) explained by the properties of phase folding partial correlations (Sect.\,\ref{sec:disc:subharmonics}). Polaris exhibited the strongest such signal at $P \gtrsim 20$\,d ($f \lesssim 0.05\, \mathrm{d^{-1}} \approx 0.6\,\mu$Hz).
                If interpreted as a real signal, then a global low-frequency power increase could be related to granulation noise. However, two studies that previously investigated such effects reported very different (and similar) timescales. \citet{Derekas2017} analysed the nearly uninterrupted high-precision 4-year-long light curve of V1154\,Cygni collected by the {\it Kepler} spacecraft and reported a detection of granulation noise on an effective timescale of $3.5 \pm 0.5$\,d, noting that this timescale agreed rather well with expectations based on the scaled properties of red giant stars. \citet{Bruntt2008} detected a power increase at low frequencies for Polaris on a timescale of $2-6$\,d using observations of the star tracker on the WIRE satellite collected during three week-long runs at high cadence. This timescale was considered an (intriguingly) good fit to a Solar granulation signal scaled to the properties of a Cepheid. Since a single pixel of the WIRE star tracker subtends one minute of arc on the sky \citep{Buzasi2002WIRE}, Polaris' known companions \citep{Evans2024}, Aa (at $<0.1"$) and B ($8$th mag, at $18"$), could have influenced the WIRE star tracker photometry at the mmag level. 
                Our data set spanning 290 observations over $6$\,yrs would suggest a significantly slower timescale, which would result in poorer agreement with the aforementioned scalings based on the Sun or red giants. Further high-cadence and long-term monitoring would be needed to settle this conundrum. For the time being, we consider the power increase to most likely be an artefact of the periodograms used here.
    
            \subsubsection{Long-period signals}
                Polaris was the only Cepheid that credibly exhibited line shape signals at periods much longer than the radial mode. Several promising signals were initially identified, but only a single signal at $59.86$\,d passed the local S/N threshold for detection after taking into account the rise in power at low frequencies. This signal was previously identified in \citet{Anderson2019} based on a data set collected with the same spectrograph, yet with minimal overlap in data. It is accompanied by a subharmonic at $119$\,d, which was previously detected using RV data by \citet{Lee2008}. Interestingly, another signal at $40.2$\,d identified by both \citet{Hatzes2000} and \citet{Anderson2019} was not recovered here, as was the case by other previous studies \citep[e.g.][]{Bruntt2008,Lee2008}. As previously reported by \citet{Dinshaw1989}, this signal may be not coherent and could vary in time. Several not-quite-significant signals found in our data set suggest that such non-coherent signals are present on timescales of $27 - 48$\,d. Longer-term monitoring may be required to ascertain whether the $\sim 60$\,d signal persists, or whether it, too, changes or vanishes with time. However, an interesting possible explanation for this signal could  be related to rotation, since Polaris has been shown to exhibit both spots \citep{Evans2024} and surface magnetic field \citep{Barron2022}. Polaris' small temperature variations may facilitate the formation of more substantial and long-lived spots, which may help explain why analogous signals have not yet been found in other Cepheids.

            \subsubsection{Line splitting signals}
                The visually striking line splitting signals of X Sgr and BG Cru, cf. Fig.\,\ref{fig:line_splitting}, have been shown to vary on a different timescale than the main radial mode \citep{anderson,Netzel2024b}. In both stars, strong additional signals are present, and we have identified the frequencies $f_{\rm X}^{\rm XSgr} = 0.0812\,\mathrm{d^{-1}}$ and $f_{\rm X}^{\rm BGCru} = 0.3325\,\mathrm{d^{-1}}$ as their independent additional signals, respectively. In BG Cru, $f_{\rm X}$ dominated the shape periodograms, whereas the combination frequency $f_1 = f_{\rm FU} + f_X = 0.2238 \,\mathrm{d^{-1}}$ dominates in X Sgr. Conversely, BG Cru's analogous combination frequency $f_{7} = f_{\rm 1O} + f_X = 0.6316 \,\mathrm{d^{-1}}$ was found in all but the Gaussian fits data set, that is where only the radial mode is significant. The frequency ratios relative to the main radial mode in both stars is $f_r/f_X = 0.900$ in BG Cru, and $1.756 = 1/0.569$ in X Sgr ($1.570=1/0.637$ if $f_1$ is used instead of $f_{\rm X}$); hence, this ratio is $1.951$ times larger in BG~Cru than in X~Sgr. Thus, it is rather unlikely that $f_{\rm X}$ is related to $f_r$, and we consider it a separate and independent signal.
    
                Further considerations concerning the origins of the line splitting feature are provided in \citetalias{Netzel2024b}, which targeted specifically classical Cepheids that exhibited resolved or unresolved line splitting. Suffice it here to say that several physical origins, such as shocks or circumstellar envelopes, can be excluded by the fact that the splitting pattern does not phase according to the radial mode.
    
                Non-radial pulsations may provide a viable alternative explanation, as they are known to distort and reshape spectral lines in complex ways. Additionally, several first-overtone Cepheids have been found \citep[see Fig. 5 in ][]{Netzel2024} to exhibit signals that fit an interpretation as non-radial modes of degrees $l=7$, $8$ or $9$ as predicted by \citet{2016CoKon.105...23D}. In these cases, the period ratio, comprised between $0.60-0.64$, is computed with the second harmonic of the suspected non-radial mode \citep{smolec2023}. The harmonic is however not necessarily always detected, complicating the interpretation. \citet{Netzel2021} have shown that the detection of such signals was theoretically possible in spectroscopic time series. However, the period ratios in BG Cru and X Sgr do not yield such a ratio. The closest case is X~Sgr, where the combination frequency $f_1$\hbox{---}which dominates the periodogram for most data sets\hbox{---}yields a period ratio of $0.637$ with the radial mode. However, this picture does not match the model by \citet{2016CoKon.105...23D}, which requires a harmonic of the non-radial mode to form this period ratio. Additionally, the light and RV curves identify X Sgr as a fundamental mode Cepheid, whereas the theory's predictions and prior detections apply to overtone and double-mode Cepheids that exhibit the first overtone. For our overtone Cepheid, BG~Cru, the Gaussian fits revealed a comparatively weak frequency $f \approx 1/2.03\,\mathrm{d^{-1}}$, which forms a ratio of $0.606$ with the radial mode similar to \citetalias{Netzel2024}. It is however under our threshold of a $p=0.001$ false-alarm probability but above $p=0.01$. We note its presence but do not consider it significant. Moreover, it is completely absent from the other data sets. We are not aware of any theoretical models that satisfactorily explain the observed period ratios in terms of non-radial pulsations. However, the additional signals might also be interpreted as a non-radial oscillation connected to resonances between the different radial modes \citep{Moskalik1992,Antonello1994, Kovtyukh2003,Smolec2010}. Unfortunately, the interpretation of such signals is complicated by an incomplete theoretical understanding of the interactions between radial and non-radial oscillations in classical Cepheids. The correlation between line splitting and FWHM of the CCFs, indicating a possible link to rotation, is unlikely as discussed in Sect. 4.3 of \citetalias{Netzel2024b}.

                In summary, the origin of these line splitting signals remains currently unknown, and we hope that the present study may usefully inspire additional theoretical work to explain these intriguing signals.

    \section{Summary and conclusions\label{sec:summary}}

        Semi-partial distance correlation periodograms are a powerful tool for detecting additional pulsation modes in pulsating variable stars, including those of currently unknown nature. These periodograms, applied to normalised spectra, CCFs, and CCF-derived data sets such as the median-residual and Gaussian fit-residual CCFs, have proven effective in isolating frequencies accompanied by Doppler shifts. We consider the full profile analysis more robust, as earlier studies assumed independence among CCF shape parameters.
        
        Our results show that additional signals dominate in shape periodograms. A comparison with classical Fourier analyses of line shape parameters, such as BIS, illustrates that these traditional methods identify signals at lower S/N and require pre-whitening by the radial mode for certain stars such as BG Cru.
        
        For $\delta$ Cep, no additional modes were found, consistent with previous findings that fundamental mode Cepheids in this period range exhibit stable pulsation properties. However, for other stars in the sample, including X Sgr, BG Cru, and Polaris, additional signals were detected. In X Sgr, the parent mode remains unclear due to data gaps; while $f_1$ consistently shows higher power, another mode, $f_{\rm X}$, has been identified through hump analysis presented in \citetalias{Netzel2024b}. Long-term uninterrupted monitoring is essential to resolve the periodicity of the line-splitting pattern in this star.
        
        In BG Cru, we identified a signal linked to line splitting and another (marginally) at the expected frequency for $0.61$ signals in first-overtone Cepheids, possibly indicating a non-radial mode. Intriguingly, the latter signal only appears in the Gaussian fit-based data set, suggesting that it might be the result of the Gaussian fitting process. This observation warrants further investigation.

        Polaris presents evidence of signals in the $20-40$ d period range, which appear non-coherent or non-stationary. A period of $\sim 59.86$ d emerges as a strong candidate for a link with the star’s rotation period, potentially explaining recent detections of star spots and magnetic fields but requires further monitoring.
        
        Median-subtracted CCFs ($\widetilde{C}$) emerge as particularly well-suited for detecting signals unrelated to pure Doppler shifts, even in low-amplitude pulsators. While full spectra also allow the detection of these signals, the power of the additional signals is typically higher in $\widetilde{C}$ data sets. Nevertheless, the wavelength-merged spectra yield the highest power for main radial (photometric) modes and enable the detection of additional frequencies in all modulated stars analysed.
        
        The ability to detect additional signals from CCF-based data sets significantly enhances the feasibility of such analyses on larger samples, overcoming the high S/N requirements of earlier methods. These observations, crucial for understanding Cepheid masses and line-broadening phenomena, call for further theoretical work. Such investigations could unlock an asteroseismic window for evolved intermediate-mass stars and the progenitors of neutron stars.

    \section*{Data availability}
        The shape and shift periodogram data sets, and cross-correlation functions (CCFs) for all four stars are available in electronic form at the CDS via anonymous FTP to \href{ftp://cdsarc.u-strasbg.fr}{cdsarc.u-strasbg.fr} (130.79.128.5) or via \href{http://cdsweb.u-strasbg.fr/cgi-bin/qcat?J/A+A/}{http://cdsweb.u-strasbg.fr/cgi-bin/qcat?J/A+A/}. Reduced spectra are available upon request to the second author.

    \begin{acknowledgements}
        The authors thank the anonymous referee for their comments, which helped improve the quality of the manuscript. We acknowledge all observers who contributed to collecting the VELOCE data set. We also acknowledge useful comments and discussions with Saniya Khan, Michaël Cretignier, Ludovic Rais, Gauthier Leurent, Ethan Tregidga, Felix Vecchi and Salomon Guinchard. This work was supported by the European Research Council (ERC) under the European Union’s Horizon 2020 research and innovation programme (Grant Agreement No. 947660). RIA is funded by the SNSF through an Eccellenza Professorial Fellowship, grant number PCEFP2\textunderscore194638. This work uses the publicly available \texttt{SPARTA} package written by A. Binnenfeld and S. Shahaf. A. Binnenfeld, S. Zucker, and S. Shahaf acknowledge support from the Israel Science Foundation (grant No.\ 1404/22) and the Israel Ministry of Science and Technology (grant No.\ 3-18143). The Euler telescope is funded by the Swiss National Science Foundation (SNSF). This research is based on observations made with the Mercator Telescope, operated on the island of La Palma by the Flemish Community, at the Spanish Observatorio del Roque de los Muchachos of the Instituto de Astrofísica de Canarias. HERMES is supported by the Fund for Scientific Research of Flanders (FWO), Belgium, the Research Council of K.U. Leuven, Belgium, the Fonds National de la Recherche Scientifique (F.R.S.-FNRS), Belgium, the Royal Observatory of Belgium, the Observatoire de Genève, Switzerland, and the Thüringer Landessternwarte, Tautenburg, Germany. HN acknowledges support from the European Research Council (ERC) under the European Union’s Horizon 2020 research and innovation program (grant agreement No. 951549 - UniverScale).
    \end{acknowledgements}

\bibliographystyle{aa}
\bibliography{bib}

\begin{appendix}
    \section{Aliases}
        Table \ref{period_results_b} lists the significant aliases detected in the different periodograms with respect to the same threshold used in the detection of the results in tables \ref{results_delCep}, \ref{results_bgcru}, \ref{results_xsgr}, and \ref{results_alfUMi}. Aliases were detected following Eq. (45) of \citet{Vanderplas2018} and a conservative tolerance of $1\%$ on the peak's frequency position relative to its alias. The two strongest features of the window were used. For BG Cru, X Sgr and Polaris, the two strongest features were the diurnal component at 1 d$^{-1}$ and 2 d$^{-1}$. For $\delta$ Cep, the strongest feature was the diurnal component and the second strongest was $f_W \cong 0.016$ d$^{-1}$. We note that for Polaris, no alias linked to these two window features were detected above the significance level.

    \label{appendixA}
        \begin{table}[hbtp]
        \caption{Overview of the most prominent aliases for each star.}
        \label{period_results_b}
        \begin{tabular}{@{}llll@{}}
        \toprule
        \multicolumn{4}{c}{$\delta$ Cep}                                         \\ \midrule
        No.      & Frequency [d$^{-1}$] & Period [d] & Identification             \\ \midrule
        $a_0$    & 0.1707315(2)                & 5.8571492 & $f_F - f_W$              \\ 
        $a_1$    & 0.2019469(2)                & 4.9517959  & $f_F + f_W$              \\
        $a_2$    & 0.6272367(2)                &  1.5942945  & $1-f_2$                      \\ \midrule
        \multicolumn{4}{c}{BG Cru}                                                  \\ \midrule
        
        $a_0$    & 0.22251311(3)                & 2.84258162   &   $(1-f_{\rm X})/3$ \\
        $a_1$    & 0.35179289(3)                & 2.84258162   &   $(1-f_{\rm 1O})/2$ \\
        $a_{2}$    & 0.4044671(3)                &  2.47238898  & $1-2f_{0}$ \\
        $a_3$    & 0.44502623(3)                & 2.84258162   &   $2(1-f_{\rm X})/3=2a_0$ \\
        $a_{4}$    & 0.50070001(3)                & 1.99720388  & $(1-f_{\rm LF})/2=a_{10}/2$ \\
        $a_{5}$    & 0.65095304(3)                & 1.53620912   & $(1+f_{\rm 1O})/2$  \\
        $a_{6}$    & 0.67034645(3)                & 1.49176593    & $1-f_{\rm X}$  \\
        $a_{7}$    & 0.70362726(3)                & 1.42120703    & $1-f_{\rm 1O}$ \\
        $a_{8}$    & 0.83515493(3)                & 1.19738262    & $1-f_{4}$ \\
        $a_{9}$    & 0.96947046(3)                & 1.03149094      & $1-f_{0}$ \\
        $a_{10}$    & 1.00098330(3)                & 0.99901766   & $1-f_{\rm LF}$ \\
        $a_{11}$    & 1.03601635(3)                & 0.96523573    & $1+f_{0}$ \\
        \midrule
        \multicolumn{4}{c}{X Sgr} \\ \midrule
        
        $a_0$    & 0.38815911(4)      & 2.57626314       & $(1-f_1)/2$                 \\
        $a_1$    & 0.61193616(4)      & 1.63415740        & $(1+f_1)/2$                    \\
        $a_2$    & 0.77626756(4)      & 1.28821562       & $1-f_1$                    \\
        $a_3$    & 0.85749957(4)      & 1.16618135       & $1-f_{\rm FU}$                    \\
        $a_4$    & 0.88947883(4)     & 1.12425385       & $1-f_0$                    \\
        $a_{5}$ & 0.92153591(4)      & 1.08514491       & $1-f_{\rm X}$                    \\ \midrule
        \end{tabular}
        \tablefoot{The window periodogram analysis reveals dominant signals at 1 d$^{-1}$ and 2 d$^{-1}$ for all stars, with the exception of $\delta$ Cep, where the second most prominent signal is detected at $f_W = 0.016$ d$^{-1}$.}
        \end{table}

    \section{Spectrum preprocessing}
    \label{appendixB}

        Ground-based observations are subject to the influence of the Earth's atmosphere, which introduces two types of unwanted modifications to our stellar observations. First, telluric lines cause a parasitic effect on the star's absorption lines. Second, the atmosphere introduces low-frequency variations in the spectrum, manifesting as modulations in its overall envelope.
        
        We conducted an extensive investigation into the impact of these phenomena on the performance of shape periodograms (K. Barbey, 2024). When telluric lines, even partially, enter the wavelength range used for the $S$ data set periodograms, they lead to inconsistent results. Conversely, the modulation of the spectral shape affects the significance of the periodogram peaks. While selecting a broader wavelength range increases the number of stellar lines, extending the range too far eventually reduces peak power in the periodogram.
        
        To mitigate the first issue, it is crucial to select a wavelength range free of tellurics. We tested telluric absorption correction tools based on line-by-line radiative transfer models, such as \texttt{Telfit} \citep{Telfit} and \texttt{Molecfit} \citep{Molecfit}, but they produced unsatisfactory results. Moreover, their computational cost was prohibitively high for use with long spectral time series.
        
        To address the overall shape modulation and normalise the spectrum, \texttt{SPARTA} applies a first-order Butterworth filter, targeting the low-pass instrumental response and high-pass instrumental noise. The filter's stop-band frequencies are set to $3(f_{\rm max}-f_{\rm min})^{-1}$ for the low-pass and $0.15/(2\Delta\lambda)$ for the high-pass, where $\Delta\lambda$ is the wavelength step of the spectrum.
        
\begin{figure}[hbtp]
            \centering
            \includegraphics[width=0.89\columnwidth]{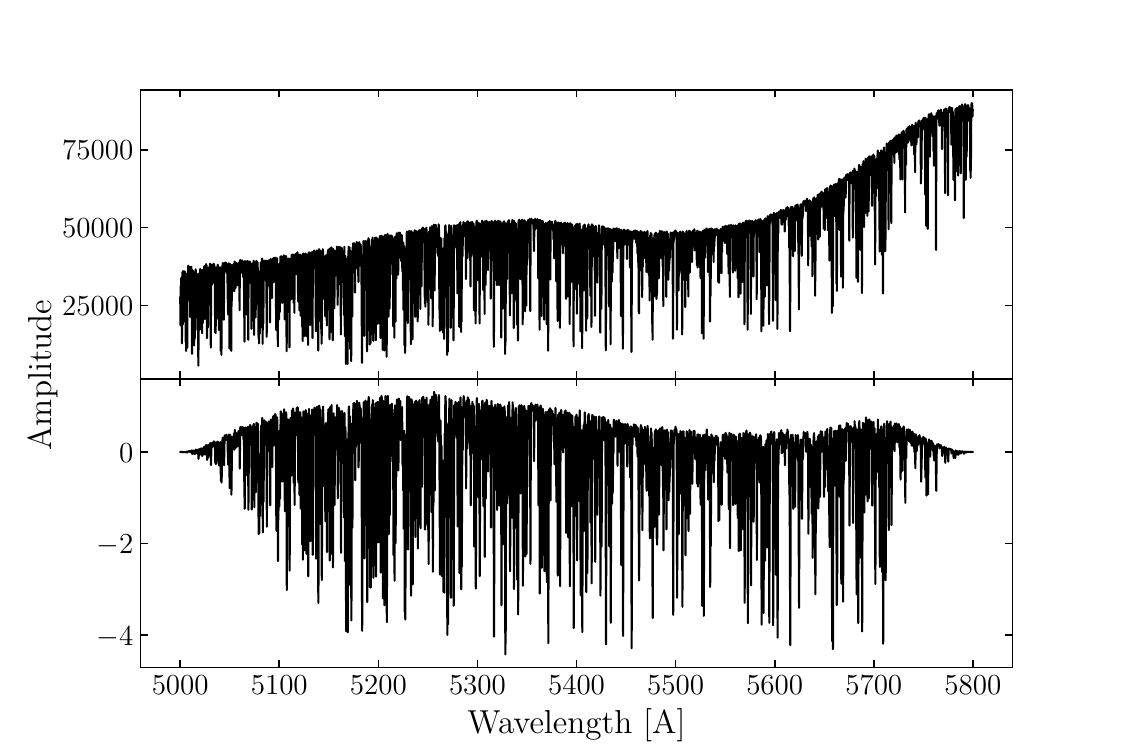}
            \caption{Comparison between the preprocessed (bottom) and non-preprocessed (top) spectra of $\delta$ Cep, observed with HERMES. The amplitudes represented on the y-axis are the flux (top) and the normalised flux (bottom).}
            \label{sparta_vs_null_HERMES}
        \end{figure}

\begin{figure}[hbtp]
            \centering
            \includegraphics[width=0.89\columnwidth]{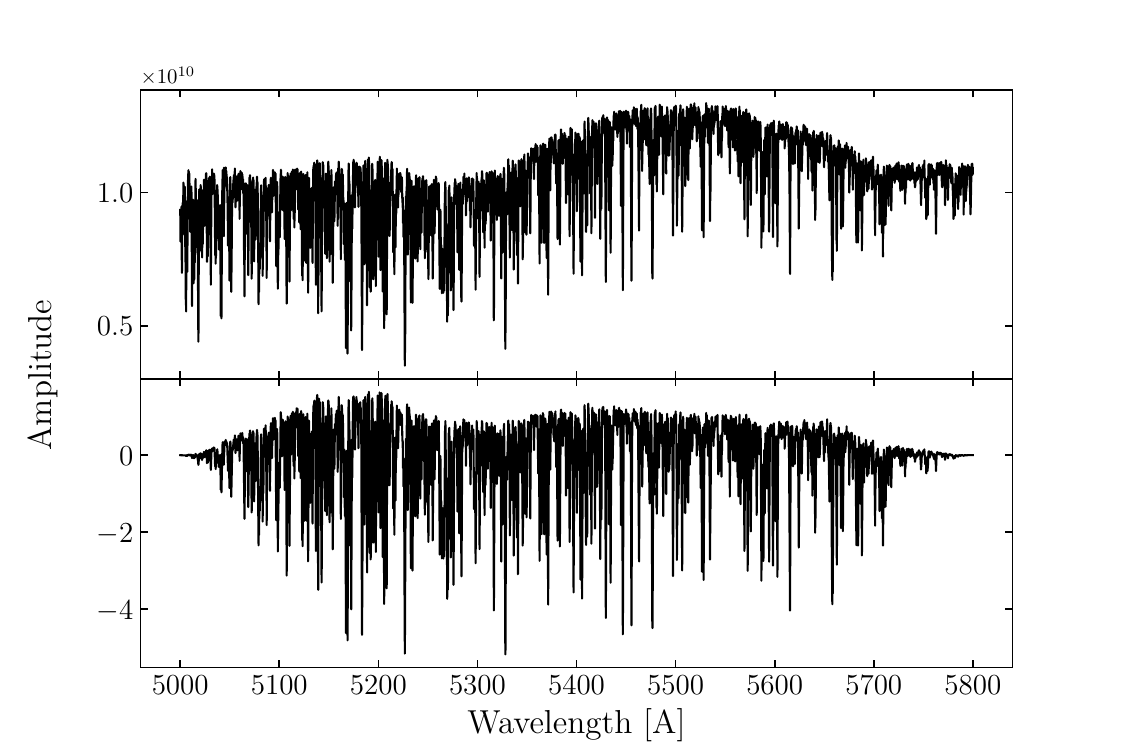}
            \caption{Same as Fig. \ref{sparta_vs_null_HERMES} but for X Sgr, observed with C14.}
            \label{sparta_vs_null_c14}
        \end{figure}

        Several other standard procedures are applied to the spectra before constructing the shape periodogram. These include interpolation, removal of cosmic rays and NaN values, and the application of a Tukey window to prevent spectral leakage. An example of the final result of this procedure is shown on Fig. \ref{sparta_vs_null_HERMES} for $\delta$ Cep observed with the HERMES spectrograph and on Fig. \ref{sparta_vs_null_c14} for X Sgr observed with C14. The wavelength range between 5000 and 5800 \AA\, offers an optimal balance in size and spectral location to effectively address both the telluric contamination and overall shape modulation issues using our available tools.

\end{appendix}
\end{document}